\documentclass[usenatbib,usegraphicx]{mn2e}

\newcommand{\chandra}{\textit{Chandra}}
\newcommand{\xmm}{\textit{XMM-Newton}}
\newcommand{\zpup}{$\zeta$~Pup}
\newcommand{\zori}{$\zeta$~Ori}
\newcommand{\tori}{$\theta^1$~Ori~C}
\newcommand{\tsco}{$\tau$~Sco}
\newcommand{\taustar}{\ensuremath{\tau_*}}
\newcommand{\Rstar}{\ensuremath{\rm R_*}}
\newcommand{\Rmin}{\ensuremath{R_{\rm min}}}

\newcommand{\vinf}{\ensuremath{v_{\infty}}}
\newcommand{\Teff}{\ensuremath{T_{\rm eff}}}
\newcommand{\lx}{\ensuremath{L_{\rm x}}}
\newcommand{\lya}{${\rm Ly}{\alpha}$}
\newcommand{\Rone}{\ensuremath{R_\mathrm{1}}}
\newcommand{\mdot}{\ensuremath{\dot{M}}}
\newcommand{\apj}{ApJ}
\newcommand{\apjs}{ApJ}
\newcommand{\apjl}{ApJ}
\newcommand{\aap}{A\&A}
\newcommand{\mnras}{MNRAS}

\begin{document}

\title[X-ray line profile modeling of $\zeta$ Ori]{Wind signatures in
  the X-ray emission line profiles of the late O supergiant $\zeta$
  Orionis}

\author[D. Cohen et al.]{David H. Cohen,$^{1}$\thanks{E-mail:
    dcohen1@swarthmore.edu} Maurice A.  Leutenegger,$^{2}$ Kevin T.
  Grizzard,$^{1,3}$ \newauthor Catherine L. Reed,$^{1}$ Roban H. Kramer$^{1,2,4}$ and  Stanley P. Owocki$^{5}$ \\
  $^{1}$Swarthmore College Department of Physics and Astronomy, 500 College Ave., Swarthmore PA 19081 \\
  $^{2}$Columbia University, Department of Physics and Columbia Astrophysics Laboratory, 550 W.  120$^{\rm th}$ St., New York NY 10027 \\
  $^{3}$St.\ John's College, 60 College Ave., Annapolis MD 21401 \\
  $^{4}$Prism Computational Sciences, 455 Science Dr., Suite 140, Madison WI 53711 \\
  $^{5}$Bartol Research Institute, University of Delaware, 217 Sharp
  Laboratory, Newark DE 19716}

\maketitle

\label{firstpage}

\begin{abstract}

  X-ray line profile analysis has proved to be the most direct
  diagnostic of the kinematics and spatial distribution of the very
  hot plasma around O stars.  The Doppler-broadened line profiles
  provide information about the velocity distribution of the hot
  plasma, while the wavelength-dependent attenuation across a line
  profile provides information about the absorption to the hot plasma,
  thus providing a strong constraint on its physical location.  In
  this paper we apply several analysis techniques to the emission
  lines in the \chandra\/ HETGS spectrum of the late-O supergiant
  \zori\ (O9.7 Ib), including the fitting of a simple line-profile
  model.  We show that there is distinct evidence for blue shifts and
  profile asymmetry, as well as broadening in the X-ray emission lines
  of \zori. These are the observational hallmarks of a wind-shock
  X-ray source, and the results for \zori\ are very similar to those
  for the earlier O star, \zpup, which we have previously shown to be
  well-fit by the same wind-shock line-profile model.  The more subtle
  effects on the line-profile morphologies in \zori, as compared to
  \zpup, are consistent with the somewhat lower density wind in this
  later O supergiant.  In both stars, the wind optical depths required
  to explain the mildly asymmetric X-ray line profiles imply
  reductions in the effective opacity of nearly an order of magnitude,
  which may be explained by some combination of mass-loss rate
  reduction and large-scale clumping, with its associated porosity-based
  effects on radiation transfer.  In the context of the recent
  reanalysis of the helium-like line intensity ratios in both \zori\ and \zpup,
  and also in light of recent work questioning the published mass-loss
  rates in OB stars, these new results indicate that the X-ray
  emission from \zori\ can be understood within the framework of the
  standard wind-shock scenario for hot stars.
\end{abstract}


\section{Introduction}

X-ray emission from normal, single OB stars has generally been
explained in terms of shock heating of the massive, highly supersonic,
radiation-driven winds of these very luminous stars
\citep*{Pallavicini1981,Corcoran1993,Hillier1993,Cassinelli1994,dhd1994,Cohen1996,Kudritzki1996,oc1999}.
The standard model involves the line-force instability, initially
noted in the context of hot star winds by \citet{ls1970} and later
investigated by
\citet*{lw1980,Lucy1982,or1984,ocr1988,Feldmeier1997a,Feldmeier1997b,ro2002}.
The wind-shock X-rays in this scenario arise naturally from non-local
radiation transfer in the context of the standard ``CAK'' line-driven
winds of massive stars \citep{cak1975}.

Other wind-shock models of X-ray production have also been proposed,
based on co-rotating interaction regions \citep{Mullan1984}, driven
shocks \citep{mc1989}, and inverse Compton scattering \citep{cw1991}.
Even in the context of the line-force instability mechanism, there are
different scenarios based on the self-excited instability
\citep{ocr1988} versus the instability seeded by perturbations at the
base of the wind \citep{Feldmeier1997b}, and one-dimensional
simulations versus two-dimensional simulations \citep{do2003}.

Despite this proliferation of models, very few observational
constraints could be put on any of these wind-shock models until
recently. This was at least partly because of the very limited data
quality of X-ray observations before the late 1990s.  Furthermore, the
idea that dynamo-driven coronal mechanisms, similar to what is seen on
the Sun, might be relevant to hot-star X-ray production, continues to
have adherents \citep{co1979,Waldron1984,Smith1993,wc2001,Smith2004}.
Indeed, models of surface magnetic field generation and dynamo
mechanisms that do not involve envelope convection, and therefore
might be applicable to massive stars, have recently been proposed
(see, e.g., \citet{mm2005} and references therein).  Speculation on
the applicability of such models to massive stars, including
specifically \zori, have, in fact, been motivated by the claims of
symmetric X-ray emission lines and anomalous X-ray line ratios in
\zori\ \citep{mw2006}.  Finally, it has recently been suggested that a
hybrid wind-magnetic shock-heating X-ray production mechanism is in
operation on at least some hot stars
\citep{Gagne1997,bm1997a,bm1997b,uo2002,Schulz2003,Gagne2005}.

The launch of the \chandra\/ and \xmm\/ telescopes in 1999, with their
high-resolution grating spectrometers, vastly improved the quality of
X-ray spectra available from OB stars.  Although these missions
provided a huge increase in the amount of information in the X-ray
data, they have not led to a consensus in the community regarding the
actual X-ray production mechanism in hot stars.  This is partly
because of the diverse behavior seen in the half-dozen or so normal
(not interacting binary) O stars thus far observed. And it is also
partly due to the lack of connection between the diagnostics usually
employed in the analysis of grating spectra of O stars and any
quantitative physical model.

The high resolution of the new X-ray grating spectrometers provides a
powerful diagnostic of plasma kinematics and location (via the effects
of continuum absorption across a line), and thus potentially a
discriminant among the various physical models, in the form of
resolved emission line profiles.  Hot plasma embedded in a fast
stellar wind produces Doppler-broadened emission lines, with the
velocity and density structure dictating the detailed form of these
profiles.  Continuum absorption by the unshocked wind is stronger on
the red sides of emission lines, as the red-shifted photons
originating in the far side of the wind traverse a larger column of
material than those emitted from the front, blue-shifted side.
Overall, then, a wind-shock mechanism, in which the shock-heated
plasma is embedded in a more-or-less spherically symmetric cold wind,
should lead to broadened and asymmetric lines with a blue-shifted
centroid and a characteristic shape
\citep{MacFarlane1991,Ignace2001,oc2001}.

The diverse behavior observed in the first hot stars observed with
\chandra\/ and \xmm\/ includes line profiles that are broad, shifted,
and skewed in the earliest O stars \citep{Kahn2001,Cassinelli2001} but
which are quite narrow in early B stars \citep{Cohen2003,Mewe2003} and
young O stars \citep{Schulz2003,Gagne2005}, with the X-ray emission
lines of late O supergiants, including \zori, having a more
intermediate morphology \citep{wc2001,Miller2002}.

The O4 star \zpup\/ seems to be generally accepted as fitting the
wind-shock paradigm, based on its line profiles.  The broad, shifted,
and asymmetric profiles are qualitatively what is expected from a
spherically symmetric wind source \citep{Cassinelli2001}.
Quantitative analysis \citep*{kco2003} shows that the hot plasma is
distributed throughout the wind above some minimum radius of emission
that is approximately half a stellar radius above the photosphere;
that it is distributed roughly as the density-squared of the bulk
wind; and that the kinematics of the hot plasma are consistent with
the underlying beta-velocity law \citep{lc1999} of the bulk wind.
This same analysis does, however, show that there is significantly
less continuum absorption than would be expected for a smooth,
spherically symmetric wind having a mass-loss rate consistent with UV
and H${\alpha}$ observations and wind opacity consistent with models.
This might be explained by a reduction in the mass-loss rate or by
inhomogeneities in the wind ( ``clumping'' or ``porosity''). To the
extent that the wind-shock picture is applicable to \zpup, it has
generally been supposed, however, that this very early O star is the
only hot star for which the standard wind-shock scenario can explain
the \chandra\/ observations.

The subject of this study, the late-O supergiant \zori, meanwhile, has
X-ray lines that are broad enough to be understood in the context of
the wind-shock scenario \citep{wc2001}. It was originally reported,
however, that there was no systematic trend in the Doppler shifts of
the emission lines observed with \chandra, and additionally, that the
forbidden-to-intercombination line ratio of Si\,{\sc xiii} indicates a
location so close to the photosphere that it could not be explained in
the context of wind-shock models \citep{wc2001}.  However, no
quantitative assessment has yet been made of the line profile shapes.
In this paper we quantitatively examine the shift and asymmetry in the
X-ray emission lines on \zori. We do this first by fitting Gaussians
to the strong emission lines in the \chandra\/ spectrum, and then by
performing a non-parametric analysis of the line shift and asymmetry,
and finally by applying the simple line-profile model that was
successfully used to fit the X-ray emission lines in the \chandra\/
spectrum of \zpup.  We show that the X-ray emission lines in \zori\
actually can be as well fit by standard wind-shock models as those in
\zpup, with a similar finding of lower-than-expected wind absorption.
We also discuss the results of our line-profile analysis of \zori\ in
light of a reevaluation of the forbidden-to-intercombination line
ratios that revises the earlier results to show no significant
conflict with a wind-shock origin for the X-rays
\citep{Leutenegger2006}.

In \S 2 we briefly describe the observational data and the properties
of \zori.  In \S 3 we assess the blue shifts and skewness of the line
profiles quantitatively but in a non-model-dependent way.  In \S 4 we
report on fits of an analytic, spherically symmetric wind emission and
absorption line-profile model \citep{oc2001} to nine lines in the
\chandra\/ spectrum of \zori.  In \S 5 we discuss the results of the
model fitting and their interpretation, including how these results
comport with other X-ray diagnostics, especially the helium-like
forbidden-to-intercombination line flux ratios and UV absorption line
mass-loss diagnostics.  Finally, we summarize our conclusions in \S 6.

\section{The \chandra\/ Data and Stellar Properties}

\begin{figure*}
\includegraphics[angle=90,width=80mm]{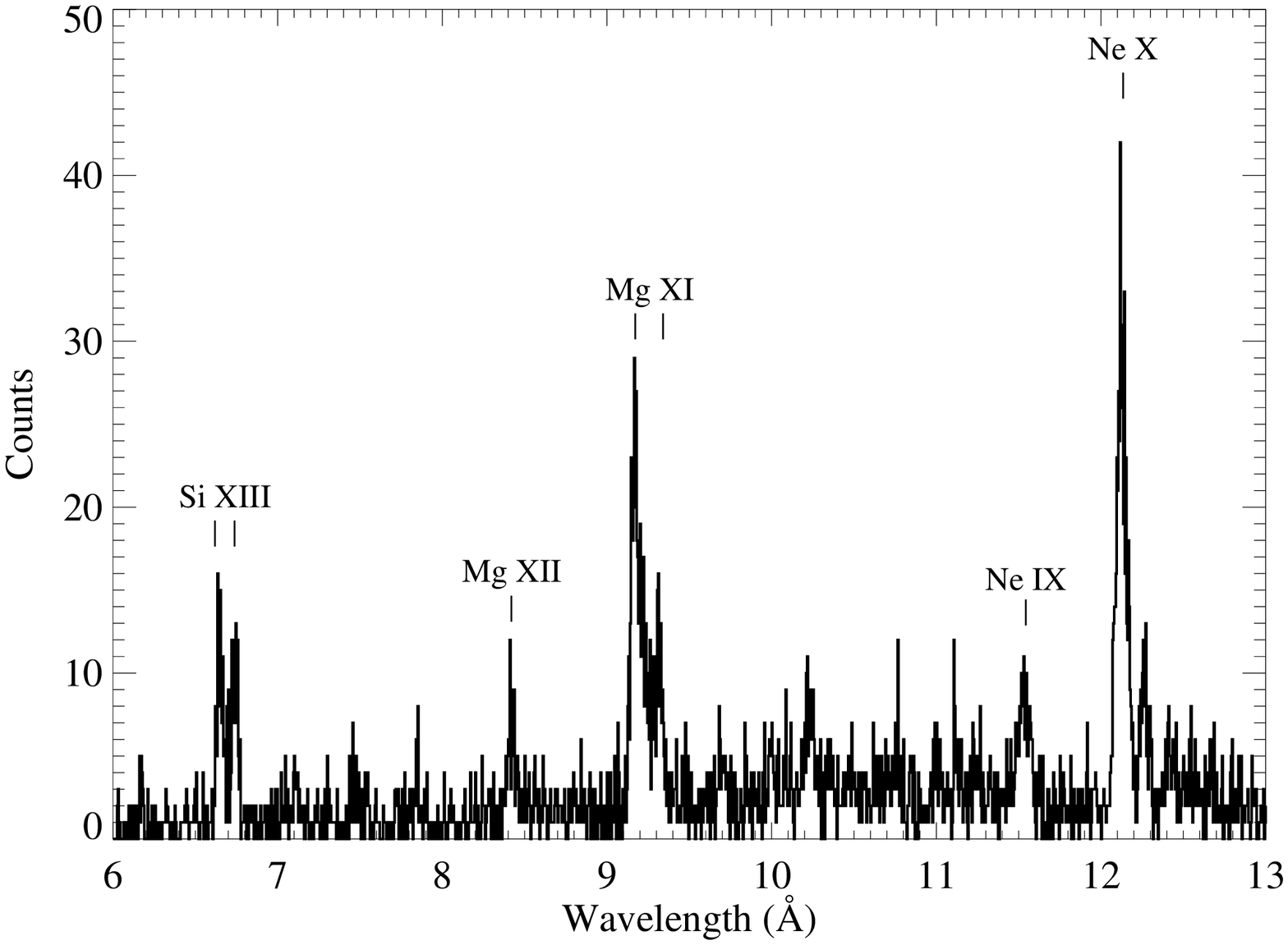}
\includegraphics[angle=90,width=80mm]{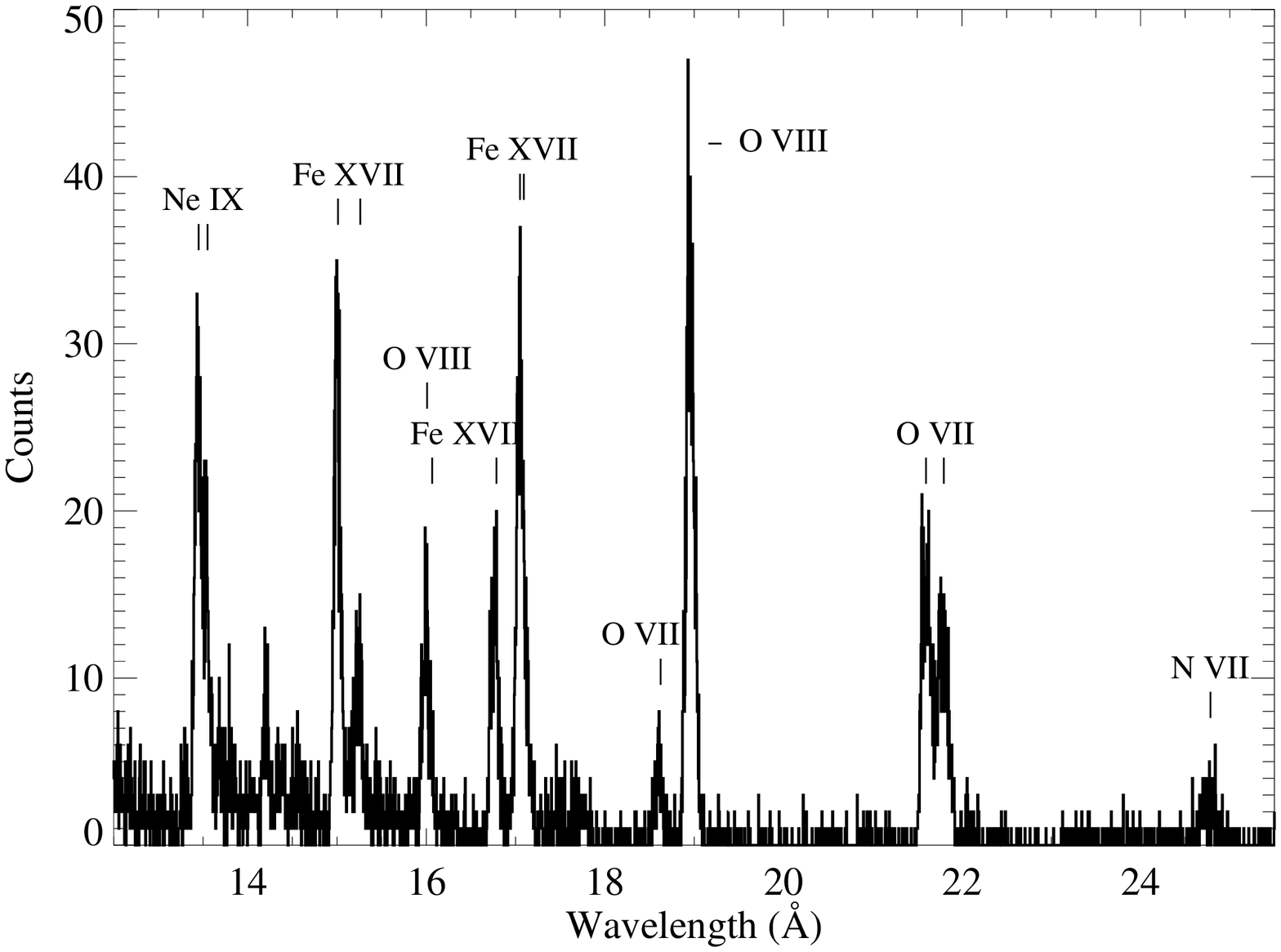}
\caption{The MEG spectrum of \zori, with negative and positive first
orders from both observations (Obs.\ IDs 610 and 1524) coadded.  }
\label{fig:atlas}
\end{figure*}

The data analyzed in this paper was obtained during the \chandra\/ AO1
GO phase, using the ACIS-S/HETGS configuration, and made with nominal
pointing at \zori.  The effective exposure time was 73.87 ks, with the
data comprising two Obs. IDs: 610 and 1524, taken on 2000 April 8 and
2000 April 9, respectively.  In the combined data, 11,347 first-order
MEG counts were recorded.  The dispersed spectrum is quite soft, as
can be seen in Figure \ref{fig:atlas}, and there were significantly
more counts in the MEG than in the HEG spectrum, which had only 2508
total first-order counts.  We therefore used only the MEG spectrum for
the line profile analysis in this paper.  We performed the standard
reduction and extraction of the dispersed spectra using the basic
grating threads and CIAO v3.1 and CALDB v2.28.  We checked the
centroids of strong lines separately in the negative and positive
first-order spectra and did not see any significant systematic shift
in the wavelengths of the emission lines between the negative and
positive sides.  We wrote the count spectra (-1 and +1 orders) to
ascii files, and performed the analysis with custom-written codes in
IDL and {\it Mathematica}, except for the initial fitting of Gaussian
line profiles, which we performed within {\it XSPEC}. We then repeated
the fits of wind-profile models to individual lines using a custom
written model within {\it XSPEC} v11.3.1.  For the {\it XSPEC}
fitting, we used only Obs.\ ID 610 (exposure time of 59.63 ks), as
including the second, much shorter, Obs.\ ID did not improve the
statistics on the fits significantly. For all the model fitting
reported on in this paper, we used the C statistic to assess the
goodness-of-fit and parameter confidence limits, as the data in the
line wings and nearby continuum have a small number of counts per bin
\citep{Cash1979}. We discuss the fitting procedure in detail in \S 4.


The late O supergiant, \zori\/ (Alnitak, HD37742, the eastern-most of
the Orion belt stars), has a Hipparcos distance of $277^{+73}_{-49}$
pc \citep{Perryman1997}. It has a spectral classification of O9.7 and
a luminosity class Ib \citep{Maiz-Apellaniz2004}, and as such is
significantly cooler than the O4 prototype \zpup, which shows X-ray
emission line profiles consistent with the wind-shock scenario.  The
wind mass-loss rate determinations for \zori\/ are roughly a factor of
two lower than those for \zpup.  Other important stellar and wind
parameters taken from the literature are listed in Table
\ref{tab:zori_params}. The overall X-ray properties of \zori\ are
quite typical of O stars (k$T_{\rm X} < 1$ keV, $L_{\rm X}/L_{\rm Bol}
\approx 10^{-7}$) \citep*{cs1983,bsc1996,wc2001}.


\section{Phenomenological and Non-Parametric Analysis of the Line Profiles}

The simplest method, and a common mode, of examining hot-star emission
line properties is the fitting of Gaussian line-profile models.  These
are convolved with the instrumental response function and multiplied
by the instrument effective area and fit to individual lines allowing
for an assessment of the centroid shifts, line widths, and amplitudes.
Indeed, this approach was taken in the paper that presented and first
discussed the high-resolution \chandra\ spectra of \zori\/
\citep{wc2001}.  \citet{wc2001} reported significant broadening
(velocity dispersion of $900 \pm 200$ km s$^{-1}$), but also noted the
generally symmetric appearance of the lines and reported a lack of any
trend in line centroid shifts.

\begin{figure*}
\includegraphics[angle=90,width=80mm]{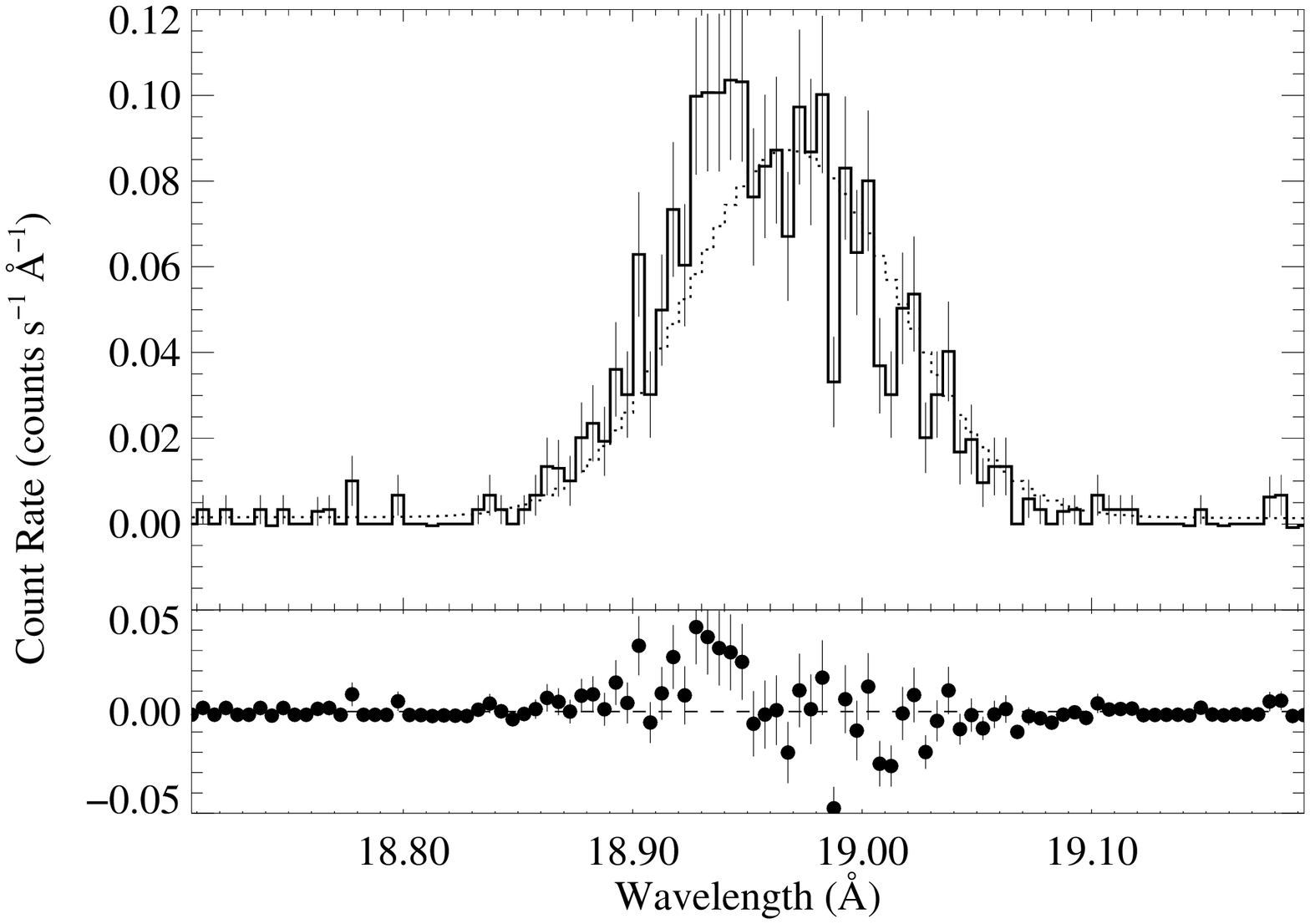}
\includegraphics[angle=90,width=80mm]{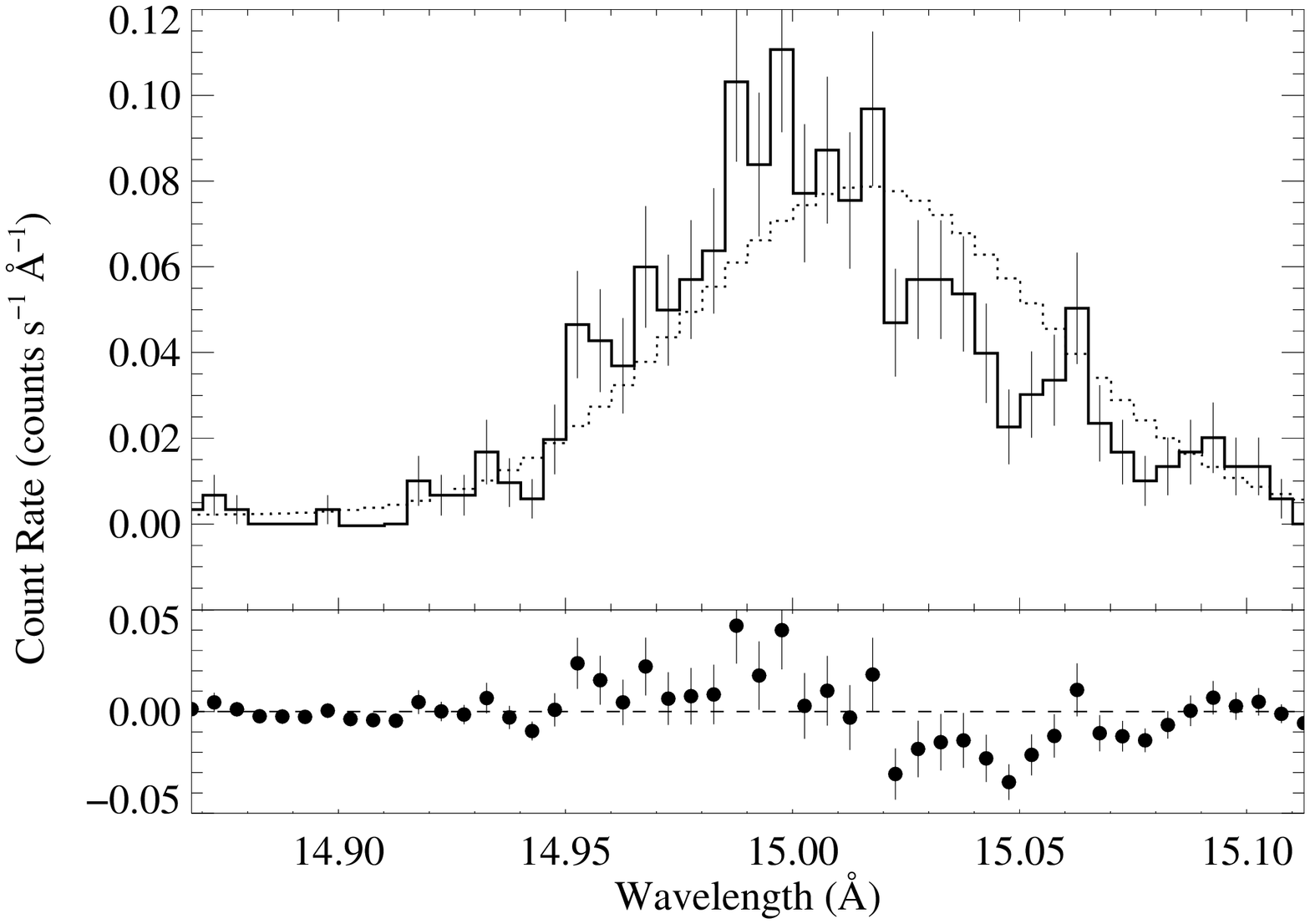}\\
\includegraphics[angle=90,width=80mm]{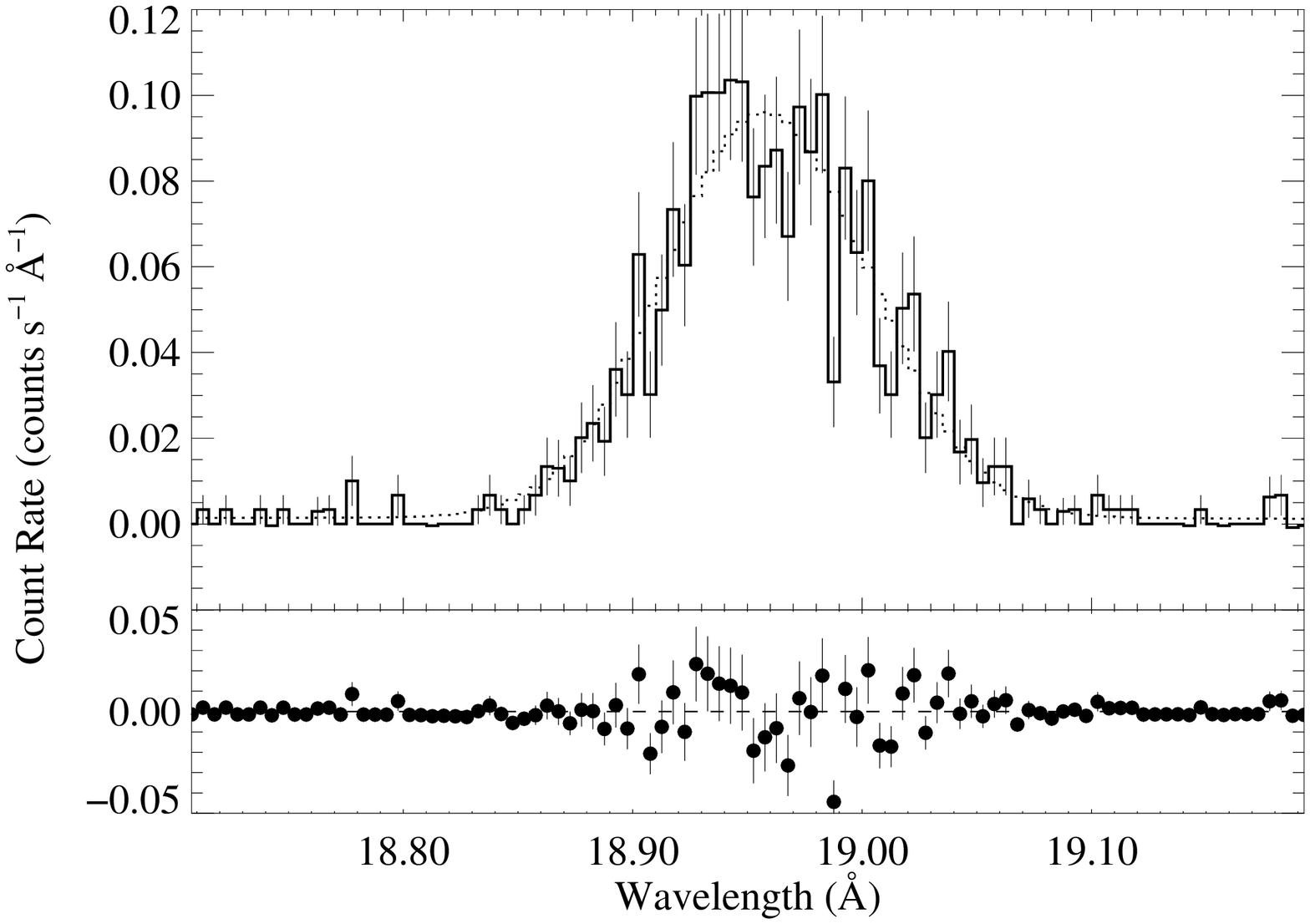}
\includegraphics[angle=90,width=80mm]{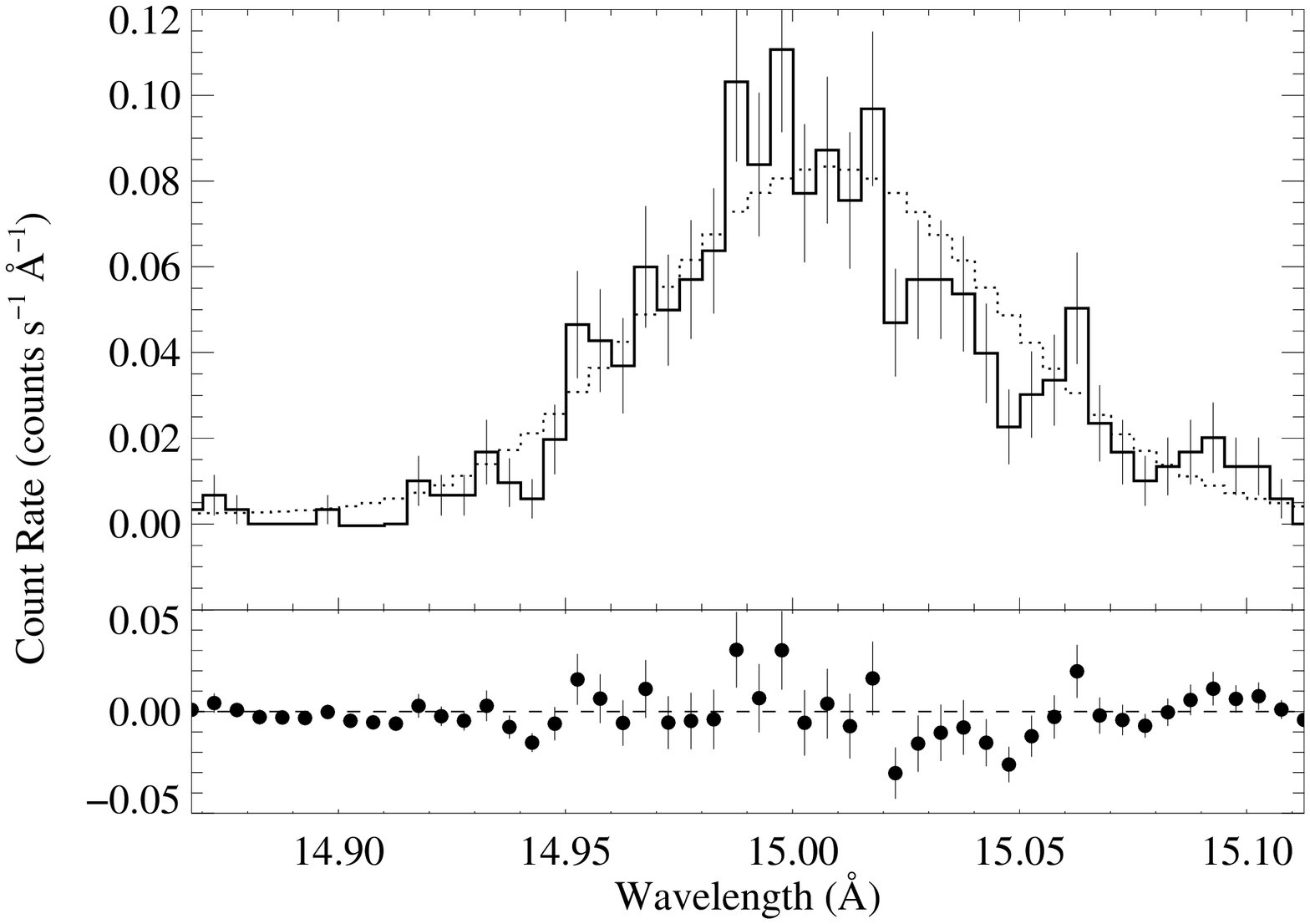}\\
\includegraphics[angle=90,width=80mm]{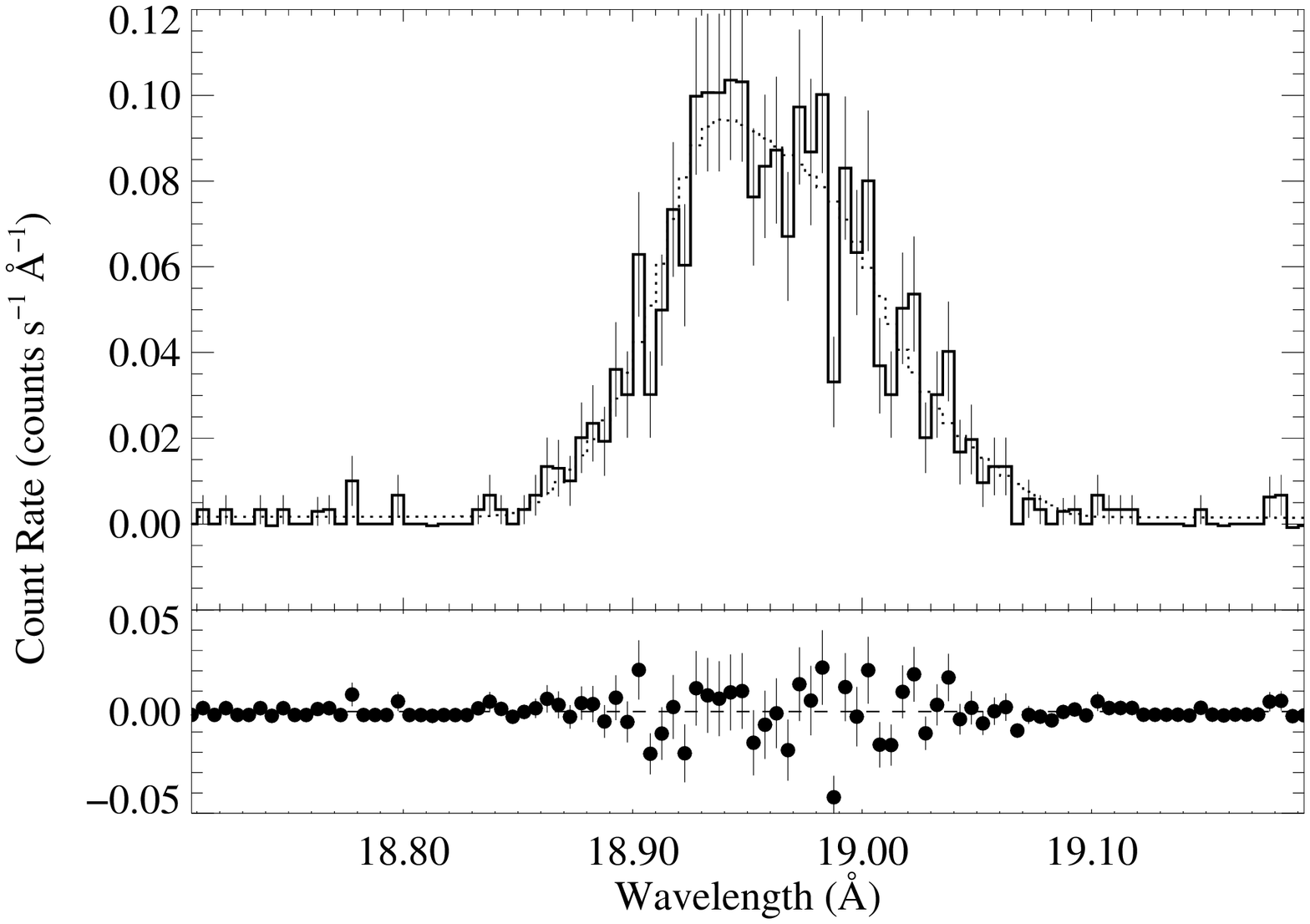}
\includegraphics[angle=90,width=80mm]{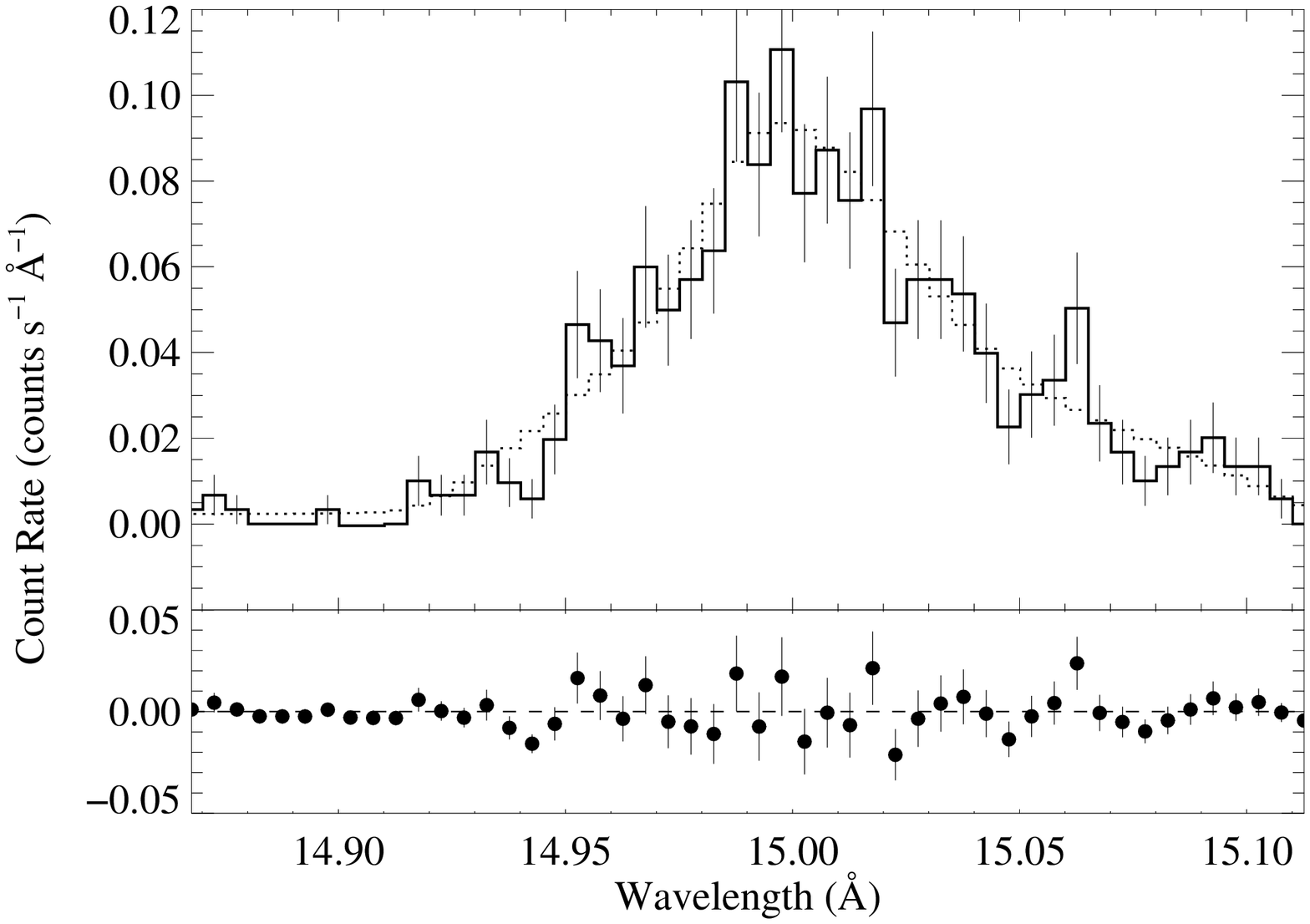}
\caption{Best-fit models superimposed on the observed O\,{\sc viii} \lya\
  line (left-hand column) and the Fe\,{\sc xvii} 15.014 \AA\ line
  (right-hand column). Error bars here and in other figures are
  calculated from the total source counts per bin, assuming Poisson
  errors.  The fits shown in the top row are for a Gaussian model with
  the line center fixed at the laboratory rest wavelength.  This fits
  shown in the middle row are for the Gaussian model with the centroid
  treated as a free parameter.  The Gaussian fits in the first two
  rows are discussed in Sec.\ 3.  The fits shown in the lower panel
  are for the wind-profile model discussed in Sec.\ 4.  }
\label{fig:gauss_fits}
\end{figure*}

We recapitulate this approach here, but also quantitatively examine
the quality of the Gaussian fits, including the distribution of the
residuals.  In Figure \ref{fig:gauss_fits} we look at two of the
strongest unblended lines in the spectrum, O\,{\sc viii} Ly${\alpha}$
at 18.969 \AA\ and Fe\,{\sc xvii} at 15.014 \AA.  In these fits, shown
in the top panels of each column, the centroid of the Gaussian was
first fixed at the laboratory rest wavelength (the
oscillator-strength-weighted mean of the two components of the \lya\
doublet in the case of the oxygen feature) and a power law was fit
simultaneously to the weak continuum.  These fits are formally bad
when analyzed using Monte Carlo simulations of the C statistic
distribution -- rejected at more than the 90\% level. There are clear
indications of line profile asymmetries in the residuals of the
Gaussian fit, in the sense one would expect from a wind-shock model,
with a blue-shifted peak and steeper blue wings and shallower red
wings.

We next fit a Gaussian model with the centroid allowed to be a free
parameter.  This model (shown in the middle panel of each column in
Figure \ref{fig:gauss_fits}) fits the line profile better, but there
are clearly still systematic trends in the distribution of fit
residuals.  Again, the actual line profiles have blue wings that are
steeper than the Gaussians and red wings that are shallower.  The
Monte Carlo analysis of the C statistic distributions shows that these
fits are better than those with the fixed Gaussian centroids, having
rejection probabilities of only 68\% and 73\%, for the O\,{\sc viii}
and Fe\,{\sc xvii} lines, respectively.

The widths and centroid shifts can be estimated from these Gaussian
fits, even if the model is not ideal.  For the oxygen \lya\ line, we
find a best fit Gaussian (half width at half maximum) HWHM of $810 \pm
30$ km s$^{-1}$, and a centroid blue shift of $-150 \pm 30$ km
s$^{-1}$.  Most other lines have even larger shifts, as can be seen in
Table \ref{tab:gauss_fits}, in which we show the results of fits to
seven emission lines in the spectrum.  These values seem plausible in
the context of the wind-shock scenario, although one might ask what
values of the peak blue shifts and HWHMs would be expected.  The
estimated terminal velocity of the wind is, after all, twice the value
of the derived HWHMs.  The answer will depend on the spatial
distribution of the X-ray emitting plasma, the velocity distribution,
and the degree of attenuation (see fig.\ 2 in \citet{oc2001}).  We
will show in the remainder of this section and the next one that there
are quantitative indications of line asymmetries, even apart from the
application of any specific wind model, and that an empirical wind
model does in fact fit the line profiles better than the shifted
Gaussian model (the wind-profile model is shown in the bottom two
panels of Figure \ref{fig:gauss_fits}, but not discussed until \S 4).
Although we cannot reject the shifted Gaussian model with a high
degree of certainty for any one emission line in the spectrum of
\zori, the Gaussian fitting suggests some degree of line profile
asymmetry and, more generally, that a more appropriate and physically
meaningful model might improve the quality of the fits.



But before fitting wind-profile models, let us first characterize the
line profile shapes using a model-independent, non-parametric
analysis. We do this by computing the first three moments of the
observed line profiles, describing respectively the centroid shift,
width, and asymmetry of the line profiles, as computed from:

\[
M_1 \equiv \frac{\sum_{i=1}^N x_i f(x_i)}{\sum_{i=1}^N f(x_i)}
\]

\[
M_2 \equiv \frac{\sum_{i=1}^N (x_i - M_1)^2 f(x_i)}{\sum_{i=1}^N f(x_i)}
\]

\[
M_3 \equiv \sum_{i=1}^N (x_i - M_1)^3 f(x_i).
\]

\noindent
Here $x$ is a dimensionless wavelength variable scaled to the terminal
velocity of the wind, with the laboratory rest wavelength of each line
set to $x=0$, as $x \equiv (\frac{\lambda}{\lambda_{\rm o}} - 1)
\frac{c}{v_{\infty}}$, and $f(x_i)$ is the number of counts in the
$i^{th}$ bin of $N$ total bins at scaled wavelength $x_i$. Note that
we have not normalized the third moment in our definition, in order to
make the calculation of its formal uncertainty more straightforward.
The standard definition of the skewness, $s$, is related to our
definition of the third moment according to 
\[
s \equiv \frac{M_3}{M_2^3
  \sum_{i=1}^N f(x_i)}.
\]

\begin{figure*}
\includegraphics[width=80mm]{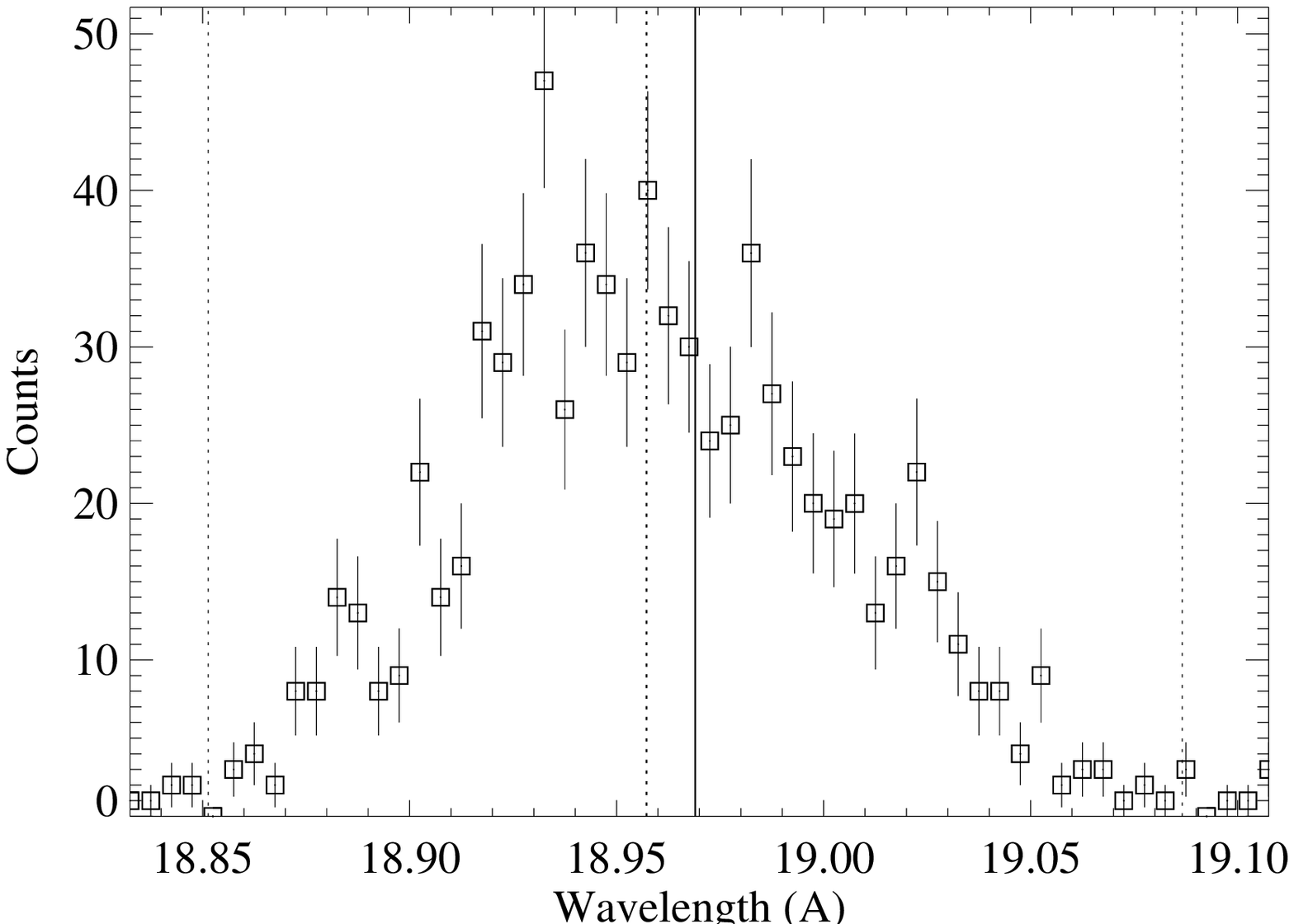}
\includegraphics[width=80mm]{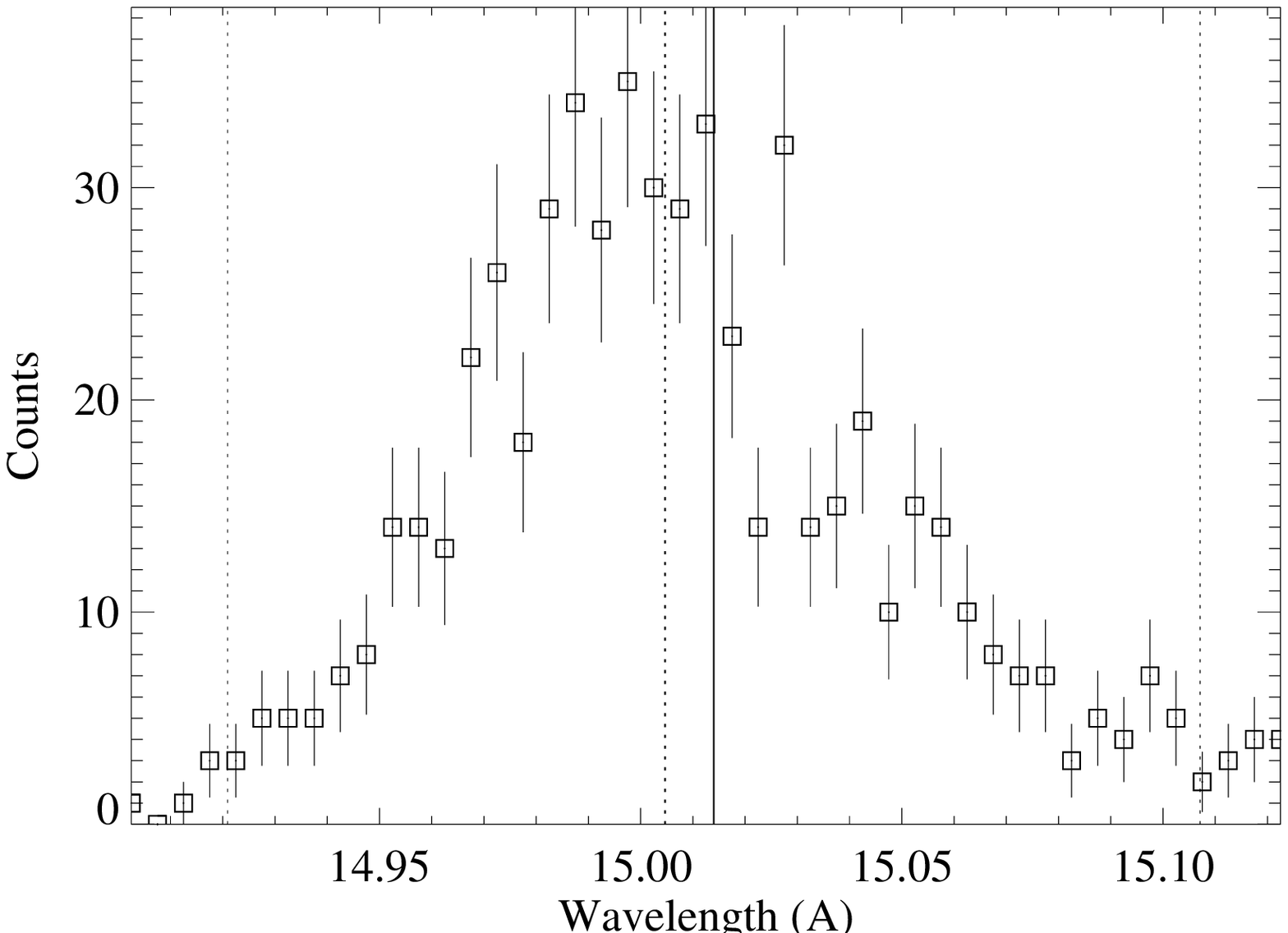}
\caption{The same two representative emission lines shown in Figure
  \ref{fig:gauss_fits}, with their centroids as determined from the
  moment analysis. O\,{\sc viii} Lyman-alpha at 18.969 \AA\ (left) has a
  centroid (first moment) 6 sigma from the laboratory rest wavelength,
  and a positive third moment (red skewed) that is significant at the
  1.7 sigma level.  The Fe\,{\sc xvii} line at 15.014 \AA\ (right) has a
  significantly negative first moment (5 sigma) and a third moment
  that is positive at the 2.3 sigma level (see Table
  \ref{tab:moments}).  In both panels, the solid vertical line is the
  laboratory rest wavelength, while the dashed line to its immediate
  left represents the first moment.  The other two dashed lines
  represent the blue and red limits over which the moment analysis was
  performed ($x=-1, 1$).}
\label{fig:moments}
\end{figure*}

We propagate the formal uncertainties for each calculated moment from
the Poisson errors on the total number of counts in each (scaled)
wavelength bin.  We note that we have not corrected for the
instrumental broadening, which is quite symmetric and not very large
compared to the observed line widths, and so will not affect the first
and third moments significantly.  We also have not corrected for the
weak continuum present under each line or for the
wavelength-dependence of the detector effective area.  But both of
these factors are explored in quantitative detail in the next section,
and are shown to be negligible.  We list the values of the first and
third moments for the stronger, unblended lines along with their
formal uncertainties in Table \ref{tab:moments}.  The second moments
are not listed, although they are quite large, because we have already
determined from the Gaussian fitting that the lines are broad and in
the moment analysis we cannot separate out the effects of physical
broadening from instrumental broadening.  We use only the unblended
lines in this analysis because the moment values have meaning only if
they are calculated on a symmetric domain about $x=0$.  In all cases
we use the domain $[-1:1]$ and assume a value of $v_\infty = 1860$ km
s$^{-1}$ for the wind terminal velocity. In Figure \ref{fig:moments}
we show two emission lines, with the moment-analysis domains
indicated, along with the laboratory rest wavelengths and the values
of the first moments.

The numerical values of the first moments are straightforward to
interpret.  They represent the position of each line centroid in units
of $x$.  The values of the third moments, however, are difficult to
interpret by themselves.  But their significance level in terms of
formal uncertainties (i.e.  their ``sigma'' levels, listed in the
final column of Table \ref{tab:moments}) are the relevant quantity for
assessing whether each line has a non-zero skewness (asymmetry) that
is statistically significant.  Unshifted and symmetric lines should
have first and third moments that are consistent with zero.  The
emission lines analyzed for \zori\/ are significantly blue shifted
(negative first moments), which is consistent with the results of the
Gaussian fitting, but which contradicts the assertion of
\citet{wc2001} that there are no systematic red shifts or blue shifts
in the emission lines.  It is also clear from the moment analysis that
the lines are significantly redward skewed (positive third moments),
generally between the 1 and 2 sigma levels for each line.  This
asymmetry was not noted in the earlier analysis, which relied on
``eyeballing'' the Gaussian fits \citep{wc2001}.  A redward skewness
(along with the blue shifted centroids) is exactly what is expected
from continuum absorption in the context of a fast, spherically
symmetric stellar wind \citep{oc2001}.  The redward skewness comes
about from the steep blue wing and the more extended, shallower red
wing.

\section{Wind Profile Model Fits to the Emission Lines}

In the previous section we showed that there is evidence for blue
shifting, redward skewness, and broadening in the X-ray emission lines
of the O supergiant \zori.  These results are consistent with the
expectations of a generic wind-shock picture. To augment this
model-independent characterization of the net profile shift and
skewness, and to derive physical information about the applicability
of a wind-shock model, let us next fit a simple, empirical wind-shock
line-profile model to the relatively strong lines in the MEG spectrum
of \zori.  We use the empirical wind-profile model of \citet{oc2001},
which is physical, in the sense that it accounts for the Doppler
shifted emission and radiation transport, including continuum
attenuation, through a three-dimensional, spherically symmetric
expanding wind.  The parameters of the model have specific, physical
meanings related to the spatial distribution of the hot plasma and the
amount of absorption by the bulk, unshocked wind.
The model is empirical, in that it does not posit any specific heating
mechanism, and thus is applicable to a wide range of possible
wind-shock (and even coronal) scenarios for X-ray emission.

The goal of fitting the wind-profile model is thus to constrain the
physical parameters of the wind emission and absorption for each
strong emission line in the \chandra\ spectrum of \zori.  Theorists
may then compare the predictions of any number of specific models or
numerical simulations to the physical parameter values we derive.
Furthermore, our fitting of wind-profile models allows us to quantify
the amount of asymmetry in the line profiles and relate the asymmetry,
quantitatively, to the amount of wind absorption, through the optical
depth parameter of the wind-profile model,

\[
\taustar \equiv  \frac{\kappa\mdot}{4\pi\vinf\Rstar},
\]

\noindent
where $\kappa$ is the absorption opacity, and \mdot\ is the mass-loss
rate. Note that bound-free continuum absorption is the dominant
opacity source. Physically, \taustar\ represents the optical depth
along a central ray from infinity to the stellar surface radius
\Rstar, in the simplified case that the wind velocity is constant at
the terminal value, $\vinf$.  In this simplified, constant-velocity
case, a value of $\taustar > 1$ also represents the radius of unit
optical depth, \Rone, expressed in units of \Rstar. The wind-profile
model assumes that the hot plasma is distributed throughout the wind,
above some minimum radius, \Rmin, and that its filling factor is
proportional to the ambient wind density multiplied by an additional
power-law factor, $f \propto r^{-q}$ (thus falling off as
$1/vr^{(2+q)}$). The other two interesting parameters of the model are
thus $\Rmin/\Rstar$ (sometimes expressed as $u_{max} \equiv
\Rstar/\Rmin$) and $q$. The normalization of the profile is the
fourth, and final, parameter.  There is an implicit assumption that
there are enough different regions of hot plasma that the wind can be
treated as a two-component fluid, comprising a bulk, cool ($T \approx
\Teff$), X-ray absorbing component, with a hot, X-ray emitting
component smoothly mixed in. The minimum radius of the hot plasma
distribution is motivated by numerical simulations that show that
large shocks tend not to form until the wind flow has reached at least
several tenths of a stellar radius
\citep{Cohen1996,Cooper1996,Feldmeier1997b}.

As discussed in further detail in \citet{oc2001}, the line profile is
computed from the integral

\[
\lx \propto \int^{\infty}_{r=r_x} \frac{r^{-(q+2)}}{(1-\Rstar/r)^{3\beta}}\exp[-\tau(\mu_x,r)]dr,
\]

\noindent
where $r_x \equiv \max[\Rmin,\Rstar/(1-|x|^{1/\beta})]$, $\mu_x \equiv
x/(1-\Rstar/r)^\beta$, and $\tau(\mu,r)$ (which is proportional to
\taustar) is the optical depth along the observer's line of sight at
direction cosine $\mu$ and radial coordinate $r$. Here the scaled
wavelength, $x \equiv (\lambda/\lambda_o - 1)(c/\vinf)$, is the same
quantity we used in the moment analysis.  The parameter $\beta$ is the
usual wind acceleration parameter, from $v = \vinf(1 -
\Rstar/r)^{\beta}$.  The governing equation for \lx\ must be solved
numerically for all $\beta \neq 0$.  We set $\beta = 1$ in all of our
fits. We include a power law continuum model in all the fits we
performed in {\it XSPEC}.  Finally, we note that this profile model
implicitly assumes spherical symmetry and a smooth wind flow.

Again, this wind-profile model is both physically meaningful and
widely applicable to a range of different physical models of X-ray
production, including coronal models (see fig.\ 2 in \citet{oc2001}
for a graphical exploration of the effects of choosing different model
parameter values on the line profile shapes, and fig.\ 4 in the same
paper for a comparison of wind-shock and coronal model parameters).
The larger \Rmin\ is and the smaller $q$ is, the broader the line
profiles tend to be.  We note that for a wide range of realistic
choices of these parameters, the characteristic width of the resulting
profiles is equivalent to roughly half the terminal velocity,
consistent with the half-widths we derived from the Gaussian fits in
the previous section.  Increasing the wind optical depth parameter,
\taustar, tends to make the profiles more narrow, more blue shifted,
and more asymmetric.  A model with a relatively small \Rmin\ value and
a negligible \taustar\ produces a profile that is similar in shape to
a Gaussian.

We fit this wind-profile model to each strong line in the \zori\ MEG
spectrum, allowing all four adjustable parameters (\taustar, \Rmin,
$q$, and the normalization) to be free, in conjunction with a
power-law component to model the weak continuum emission.  For several
line complexes, we fit multiple profiles simultaneously to account for
blending.  This included the helium-like resonance and
intercombination lines of oxygen.  We do not give fits to the other
helium-like complexes in the \zori\ \chandra\ spectrum, as fits to
these complexes are reported elsewhere \citep{Leutenegger2006}. For
the Fe\,{\sc xvii} lines at 17.051 \AA\ and 17.096 \AA\ ($3G$ and
$M2$, respectively), which we fit simultaneously, because they are
quite blended, we fixed the relative normalizations to
$I(M2)/I(3G)=0.8$, consistent with {\sc HULLAC} calculations \citep*{mlf01}
and with the values generally observed in stars. When fitting these
lines, and also other line blends, we tied the three primary
parameters of the line-profile model -- $\taustar, \Rmin, q$ --
together. Ultimately, we report here on the fits to the nine lines in
the spectrum that provide meaningful constraints to the model
parameters. Note that these are not the same set of lines to which we
fit Gaussian models, as it was easier to get meaningful Gaussian fits
to several weaker lines and it was harder to get meaningful Gaussian
fits to the blended lines.

\begin{figure*}
\includegraphics[angle=90,width=80mm]{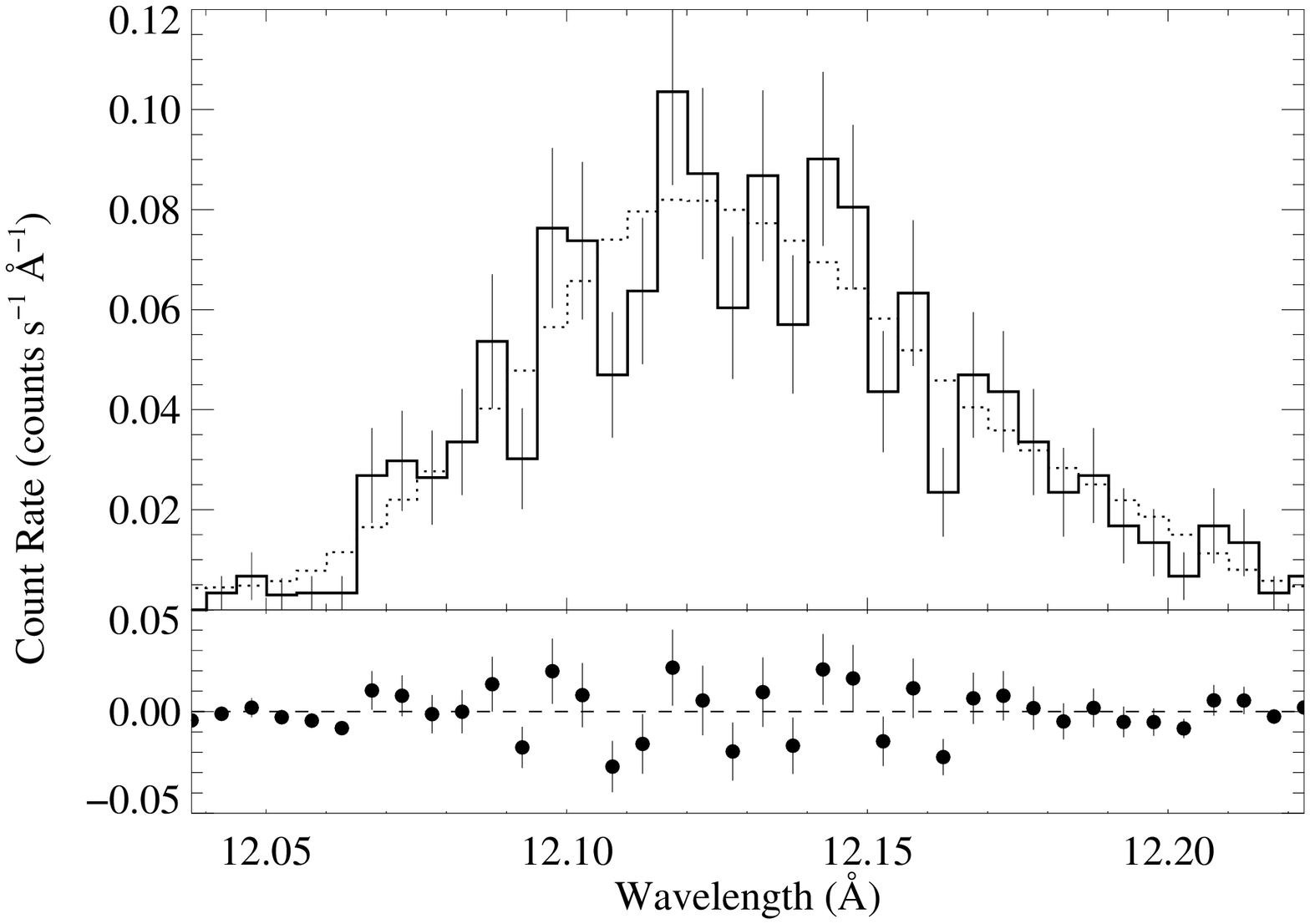}
\includegraphics[angle=90,width=80mm]{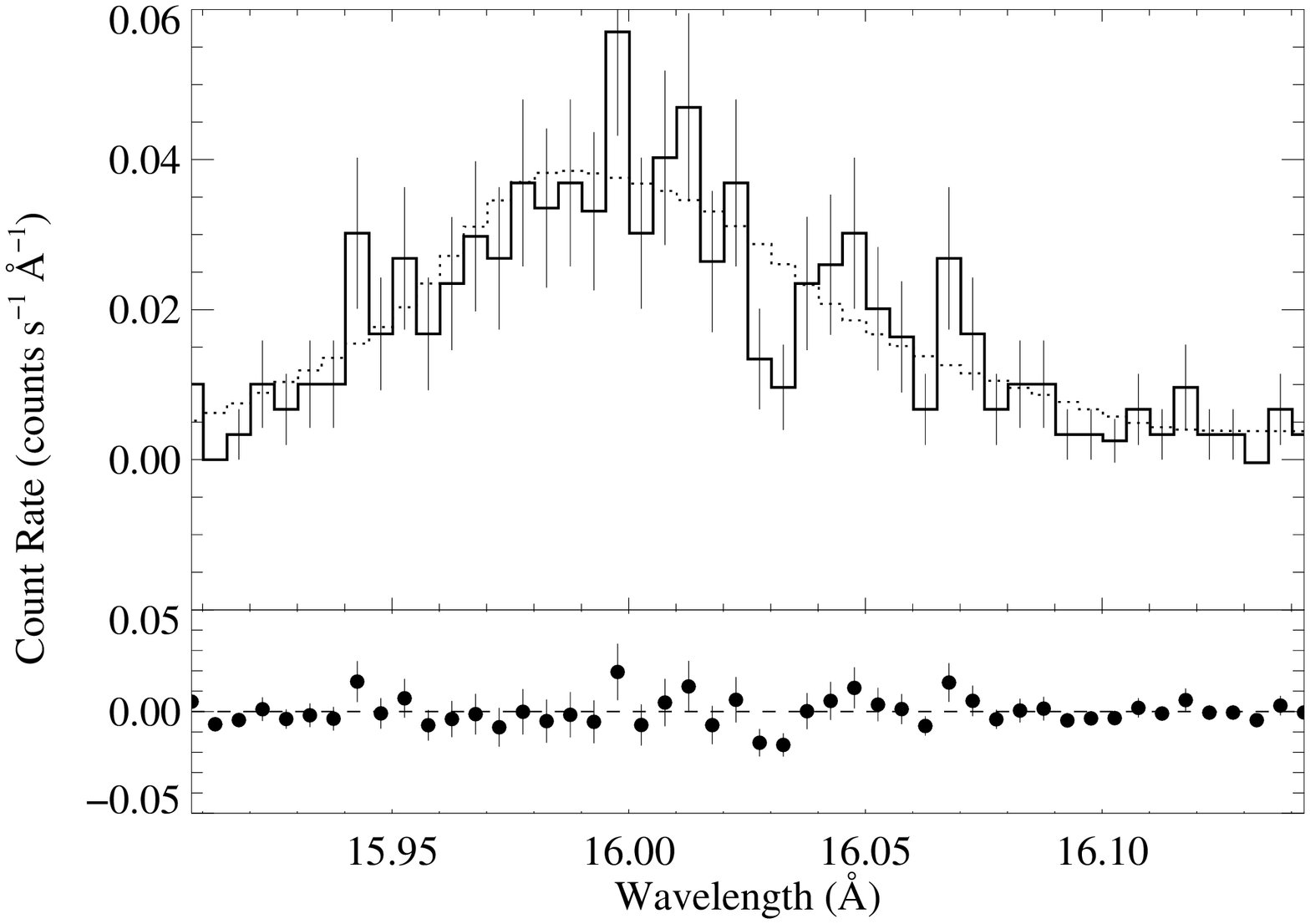}
\includegraphics[angle=90,width=80mm]{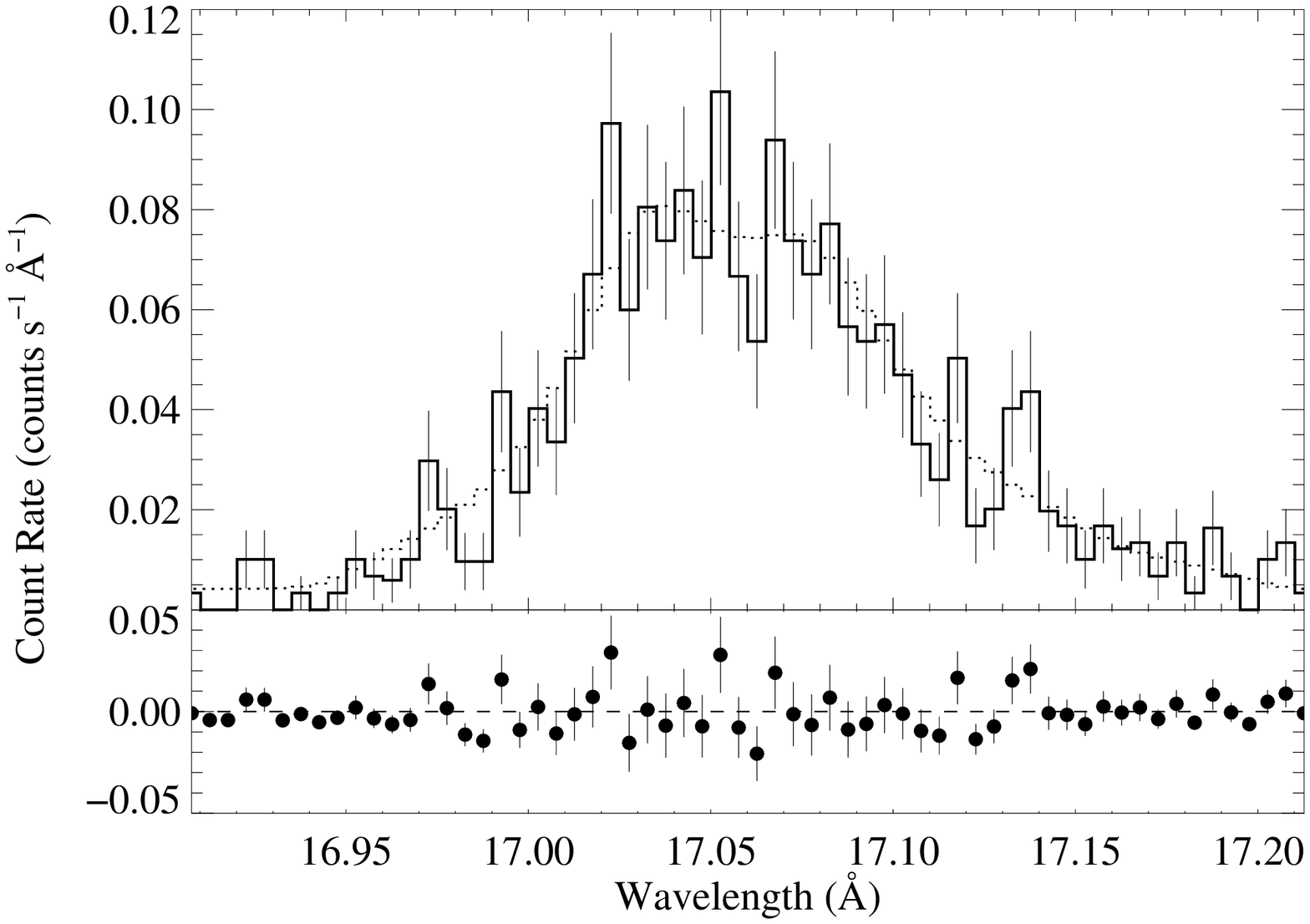}
\includegraphics[angle=90,width=80mm]{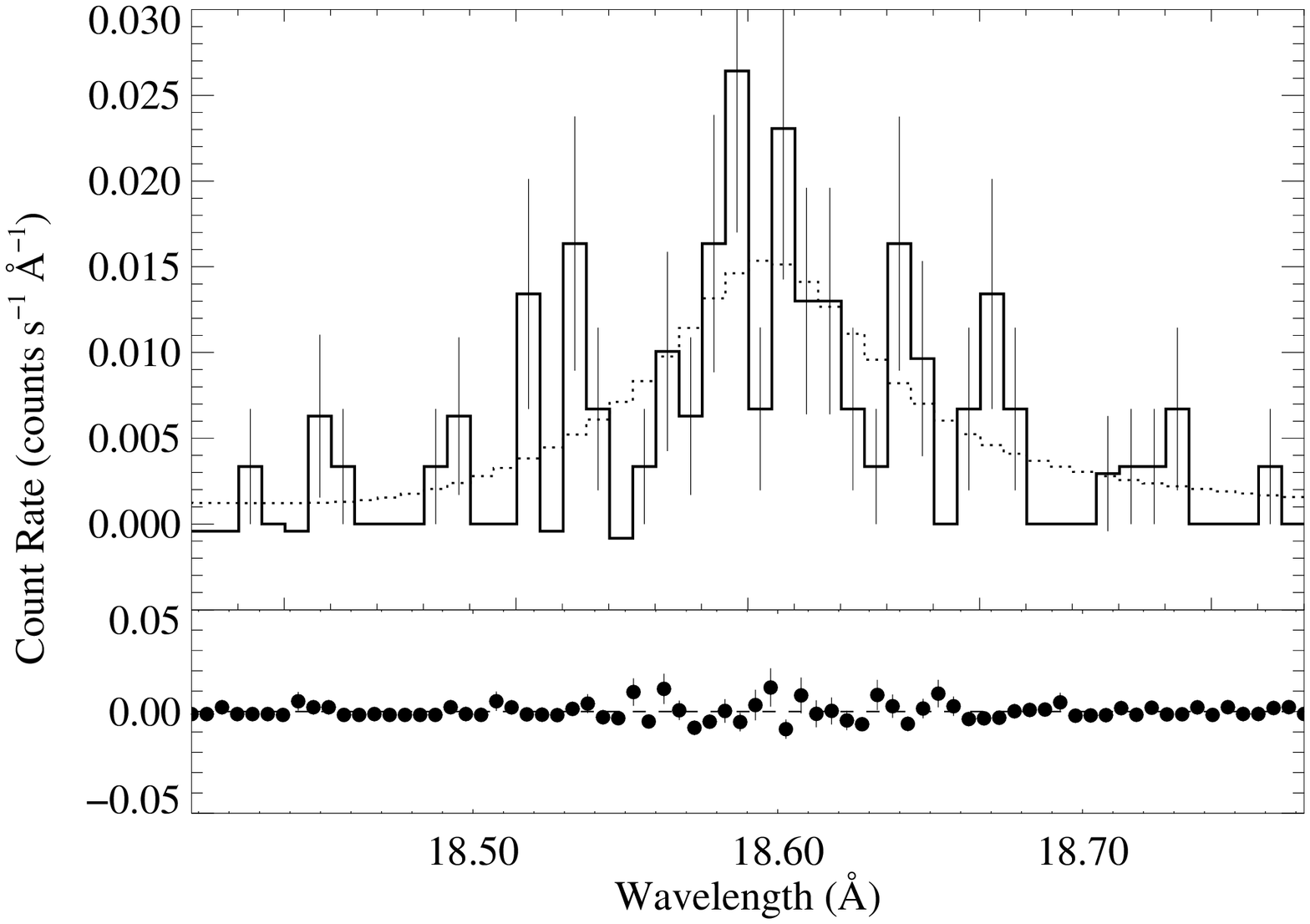}
\includegraphics[angle=90,width=80mm]{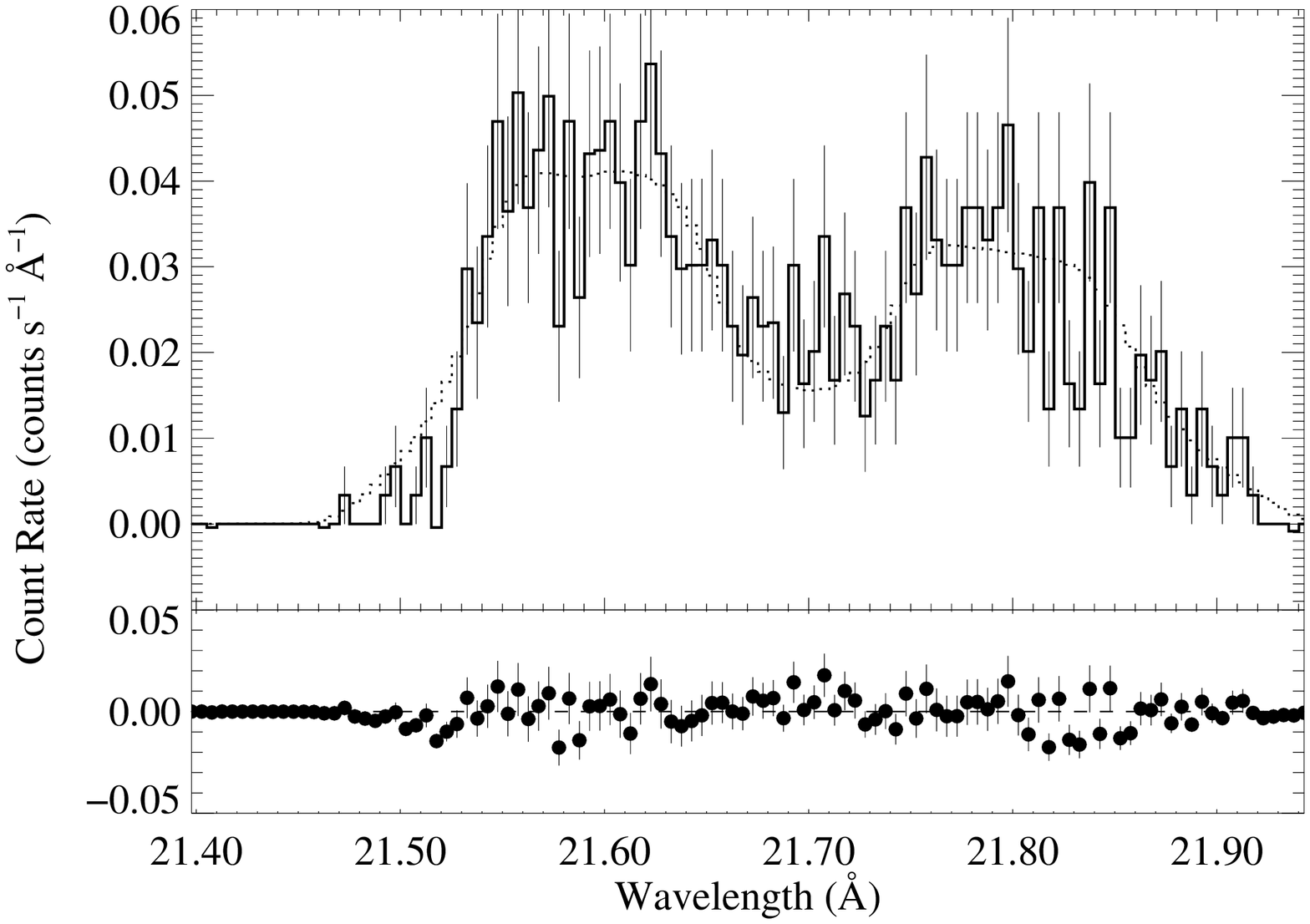}
\caption{Best-fit wind-profile models for five lines (or line
  complexes): Ne\,{\sc x} 12.134 \AA, O\,{\sc viii} 16.006 \AA, Fe\,{\sc xvii} 17.051
  \AA\ and 17.096 \AA, O\,{\sc vii} 18.627 \AA, and O\,{\sc vii} 21.602 \AA\ and 21.804 \AA. }
\label{fig:model_fits}
\end{figure*}

As mentioned previously, we first carried out this modeling using the
same procedure, implemented in {\it Mathematica}, that we employed in
our earlier analysis of \zpup\/ \citep{kco2003}.  We then repeated the
modeling using a custom-written module in {\it XSPEC}, which allowed
us to include a continuum emission component in the modeling and use
the exact instrumental responses.  This also enabled us to
simultaneously fit multiple models to line blends.  The two methods
gave very similar results, and the {\it XSPEC} fits were insensitive
to both the choice of continuum model and the wavelength range
included in the fit. The results of the {\it XSPEC} model-fitting are
summarized in Table \ref{tab:fit_params}, and the best-fit models,
superimposed on the data, are shown in Figure \ref{fig:model_fits} and
the bottom row of Figure \ref{fig:gauss_fits} for the nine lines, in
seven complexes, for which we could obtain meaningful fits. The
wind-profile model does indeed provide better fits to the stronger,
unblended lines than does the Gaussian model, according to the Monte
Carlo simulations of the distribution of C statistic values. The
goodness of fit values (expressed as a percentage of the Monte Carlo
simulations that gave a C statistic as good as or better than that
derived from the fit to the actual data; lower percentages are better)
are listed in Table \ref{tab:fit_params}.  All the wind-profile fits
are formally good.

\begin{figure*}
\includegraphics[angle=90,width=80mm]{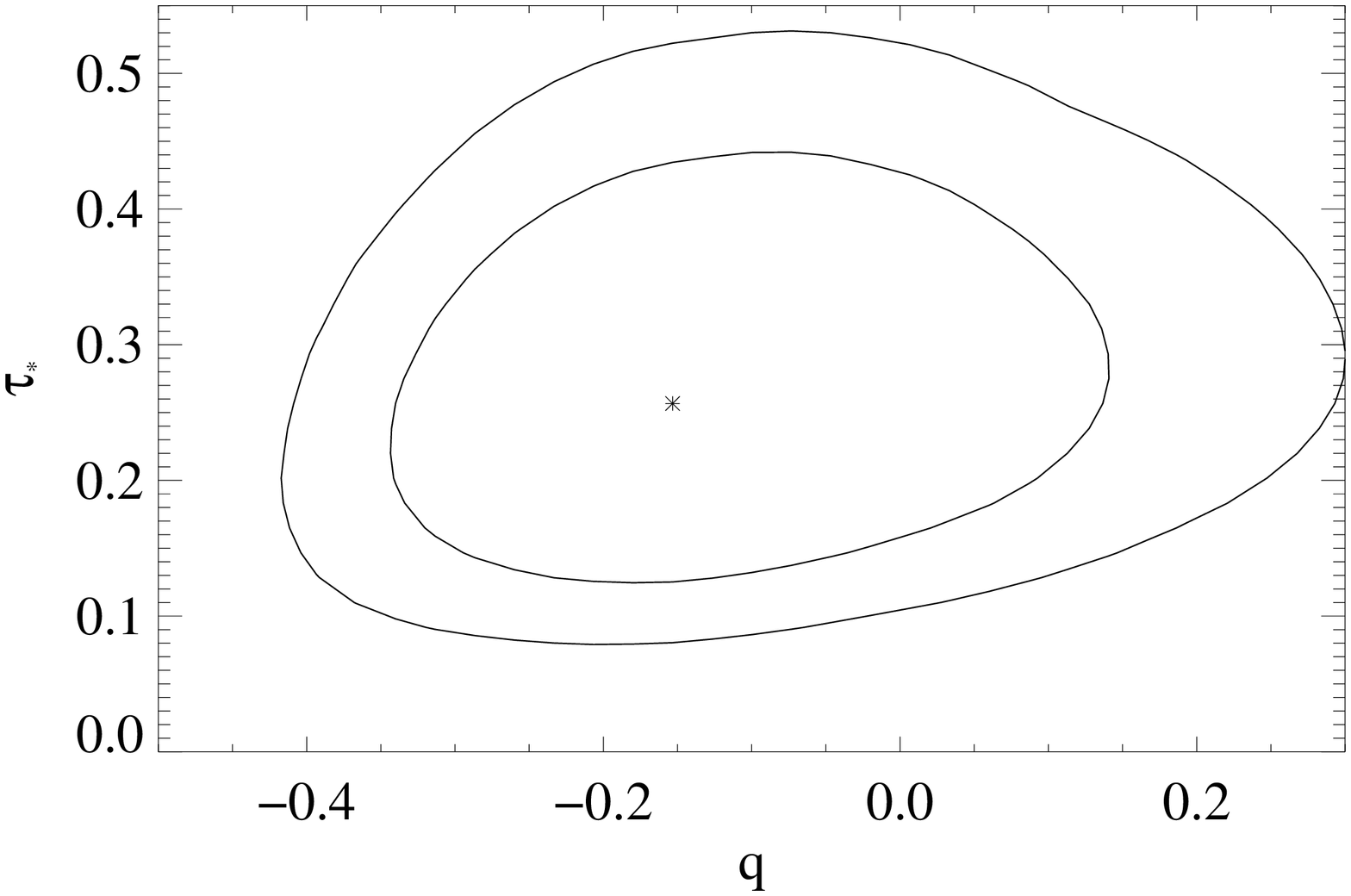}
\includegraphics[angle=90,width=80mm]{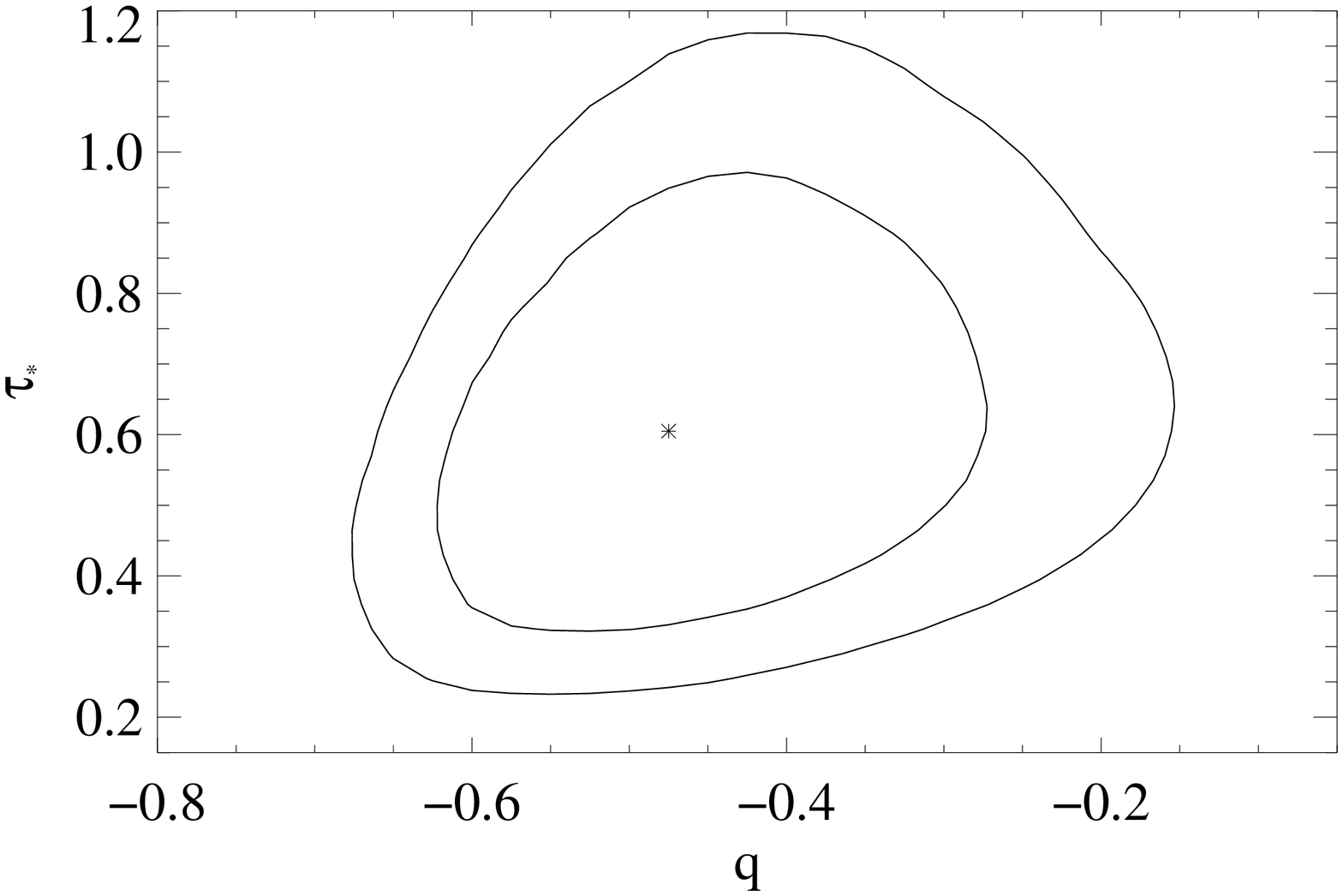}\\
\includegraphics[angle=90,width=80mm]{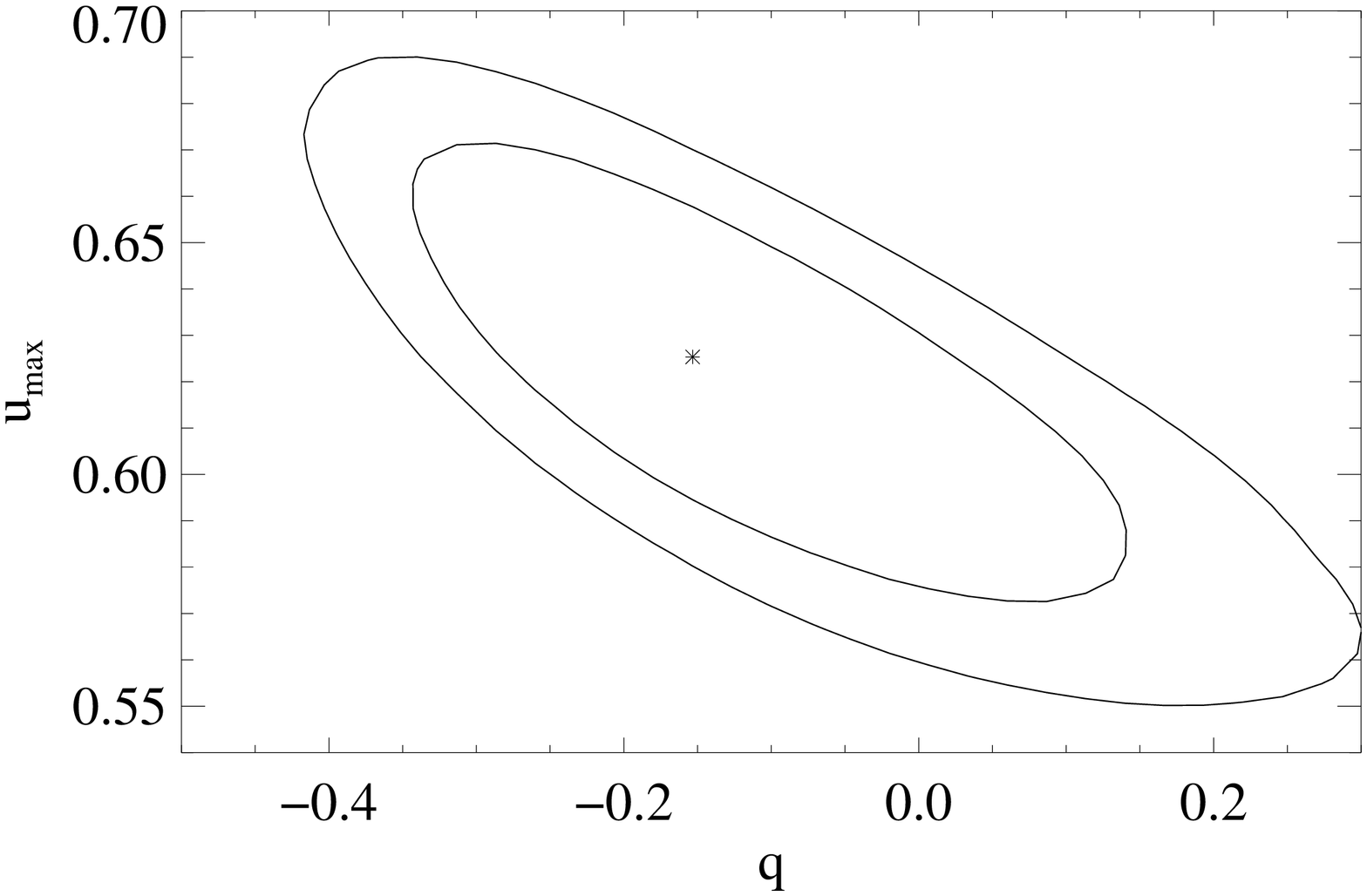}
\includegraphics[angle=90,width=80mm]{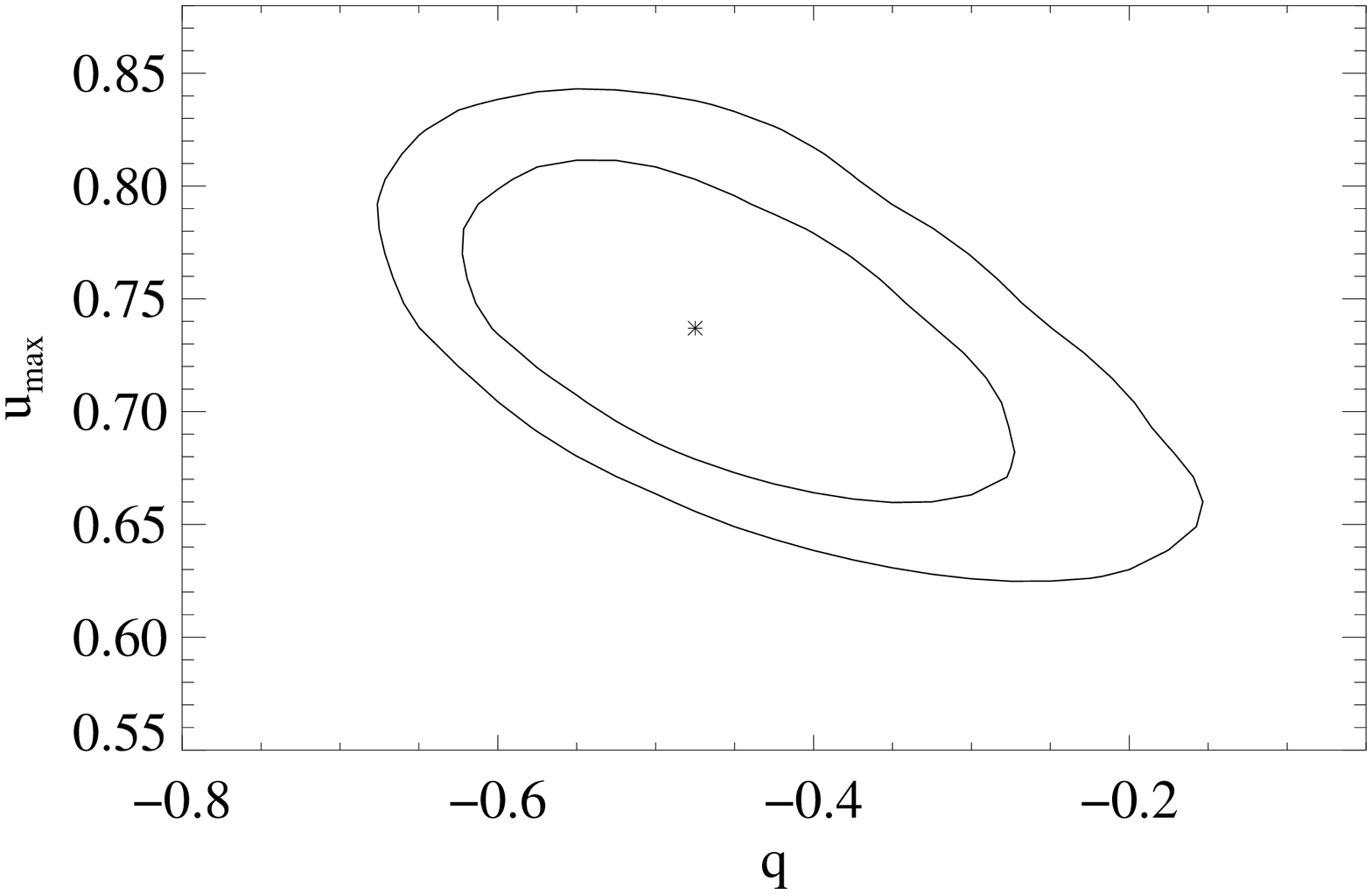}\\
\includegraphics[angle=90,width=80mm]{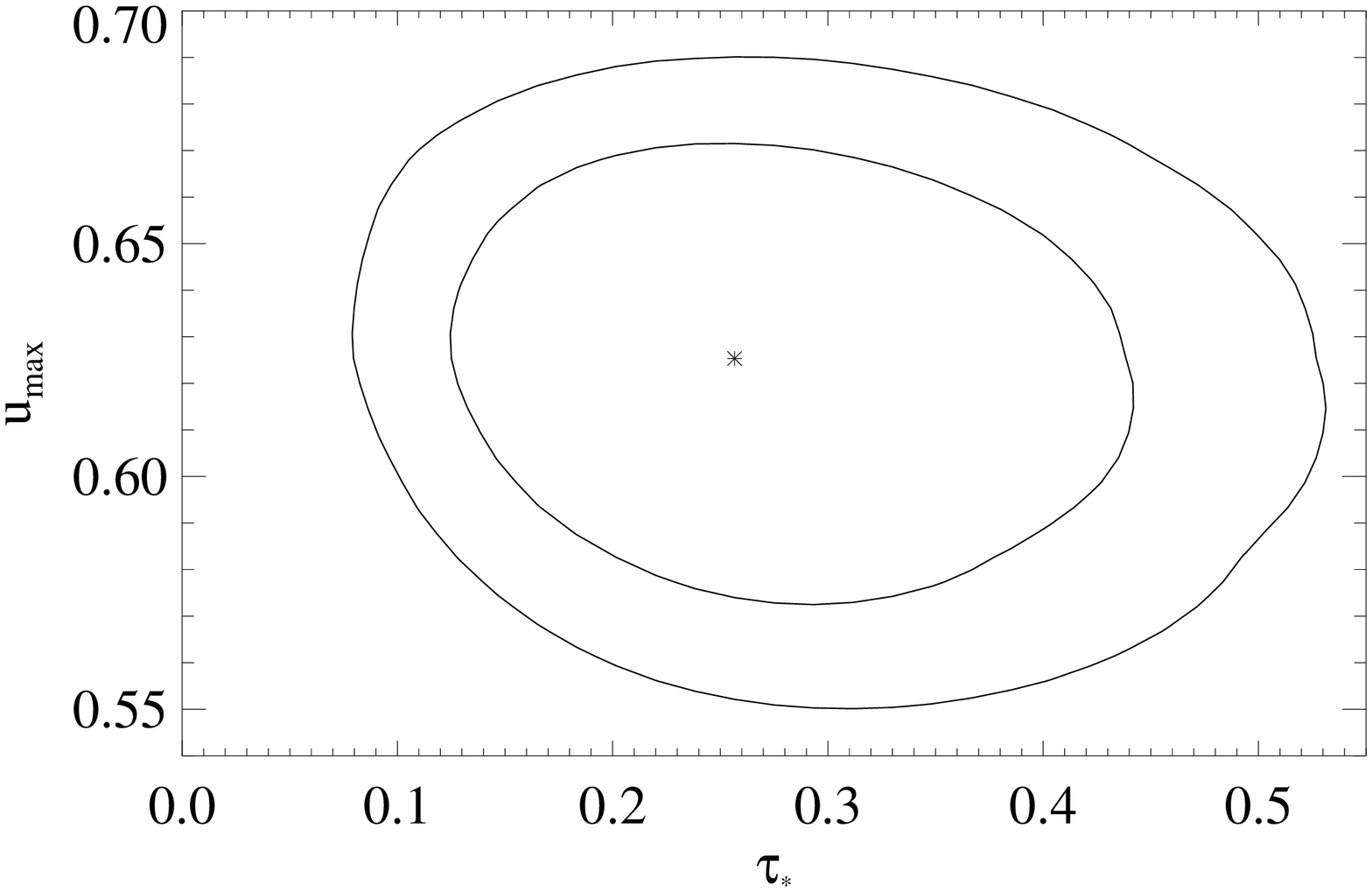}
\includegraphics[angle=90,width=80mm]{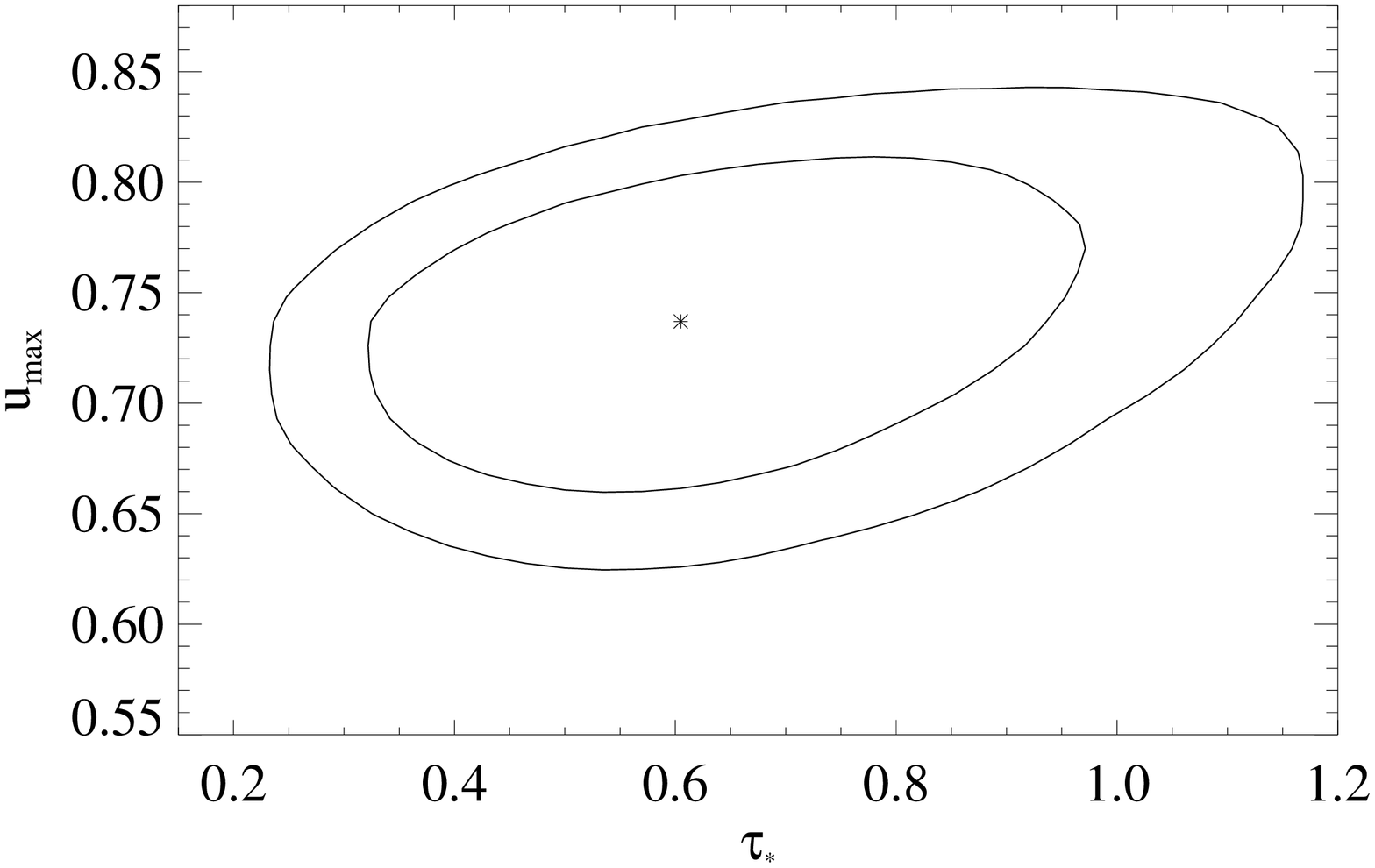}
\caption{The 68\% and 90\% confidence regions in the parameter space
  of the wind-profile model, for the two lines shown in Figure
  \ref{fig:gauss_fits}, the O\,{\sc viii} \lya\ line (left-hand column)
  and the Fe\,{\sc xvii} 15.014 \AA\ line (right-hand column). The
  best-fit model parameters are indicated by the asterisks.  Note the
  correlation between $q$ and \Rmin\ ($u_{\rm max}^{-1}$). For each
  2-D slice of parameter space shown here, the other model parameters
  are optimized (i.e.\ free) while models are fit for a grid of the two
  displayed parameter values. The contour levels thus correspond to
  $\Delta C$ values appropriate for two parameters of interest: 2.30
  and 4.61. }
\label{fig:confidence_region}
\end{figure*}

We calculated errors on the derived model parameters by using a
three-dimensional grid of models in the parameter space of interest
(\taustar -- \Rmin\ -- $q$) and applying a ${\Delta}C$ criterion
appropriate for jointly-distributed uncertainties for three
parameters, and report the maximum extent of this confidence region in
each of the three parameters as the formal uncertainties on the
derived parameters.  These are the values listed in Table
\ref{tab:fit_params} (90\% confidence limits for one parameter of
interest -- ${\Delta}C = 2.71$), and shown, for two particular fits,
in Figure \ref{fig:confidence_region}.

\section{Discussion}

\begin{figure*}
\includegraphics[width=90mm]{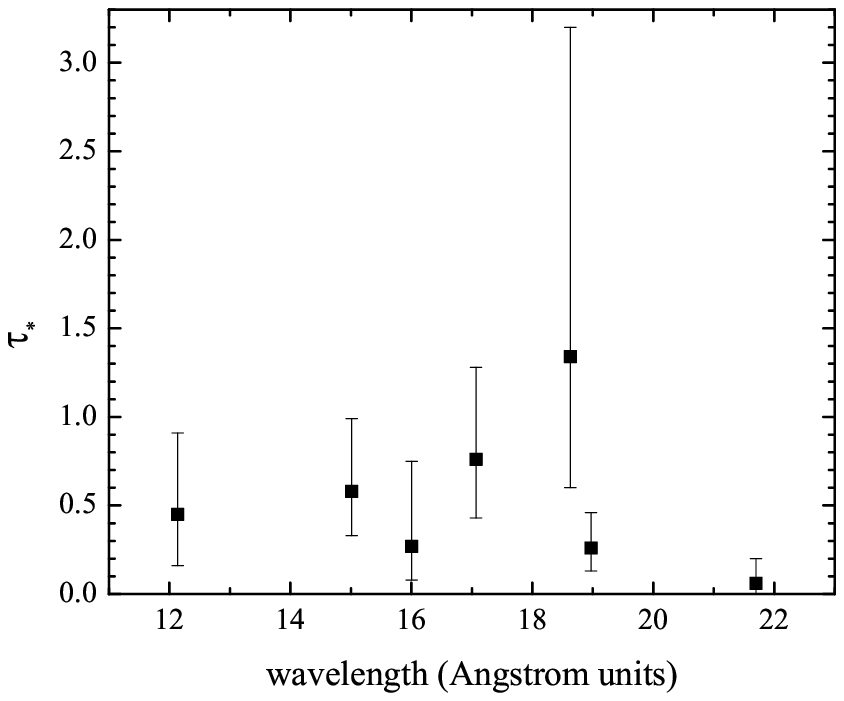}\\
\includegraphics[width=90mm]{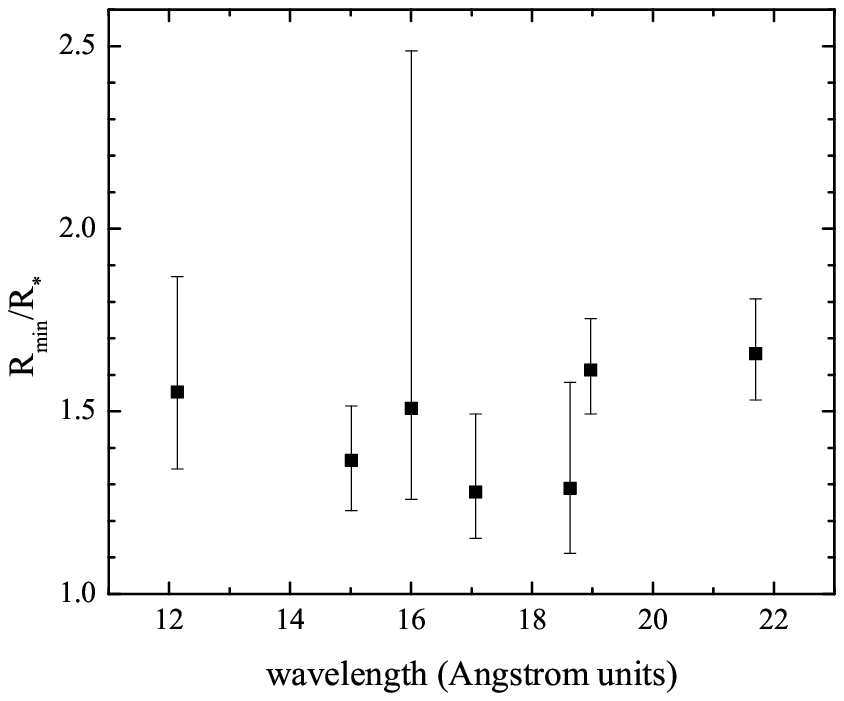}\\
\includegraphics[width=90mm]{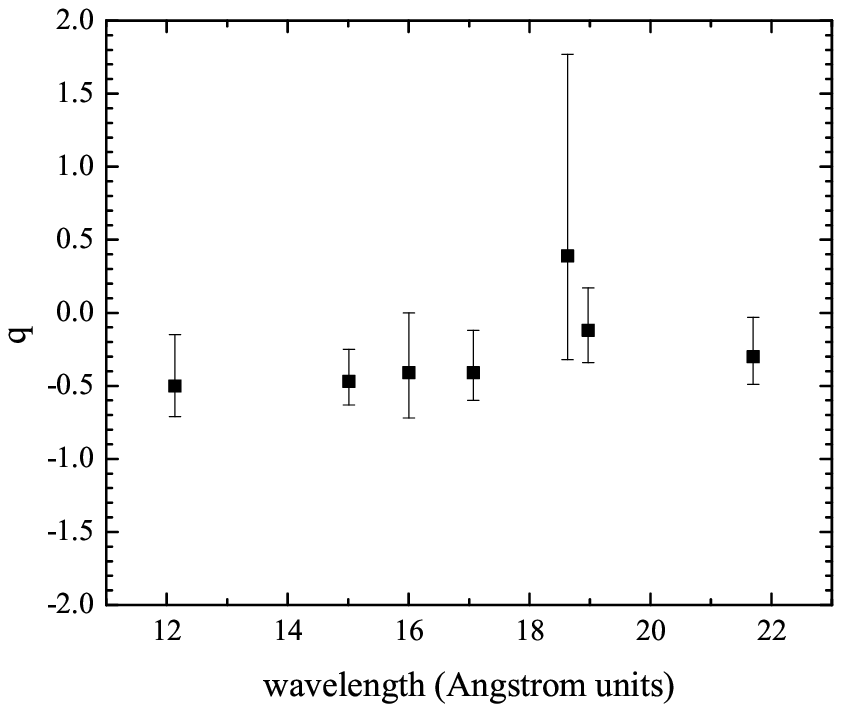}
\caption{The best-fit model parameters -- $\tau_{*}$, $q$, and $R_{\rm
    min}$ -- for each line complex we fit with a wind-profile model.
  The error bars represent the 90\% confidence limits on the one
  parameter of interest in each panel ($\Delta C = 2.71$).}
\label{fig:fit_params}
\end{figure*}

We summarize the derived model parameters and their uncertainties for
each line in Figure \ref{fig:fit_params}. This figure shows that there
are no strong trends in any of the wind-profile model parameters with
wavelength (or any other characteristic) of the emission lines.
Fitting a function linear in wavelength to the uncertainty-weighted
model parameters shows consistency with a constant function (at the
95\% confidence level) for each of the three parameters. Note that the
\taustar\ point for the O\,{\sc vii} complex near 22 \AA\ must be
excluded for this statement to be true.  We discuss this outlier in
terms of the wavelength dependence of the wind opacity near the end of
this section.

The fitting results shown in Figure \ref{fig:fit_params} present a
consistent picture of a line profile model with $\taustar \approx
0.25$ to 0.5, an onset radius, $\Rmin \approx 1.5 \Rstar$, and a
constant filling factor ($q \approx 0$).  These are all reasonable
parameters in the context of the general instability-driven wind-shock
model, though the $\taustar$ values are small compared to the
expectations of wind theory, which we elaborate on below.  Finally, we
note that most of the lines cannot be well fit by models with no wind
absorption ($\taustar = 0$ is ruled out), which is consistent with the
inability of Gaussian models to provide good fits and also with the
non-zero third moments of the line profiles, as discussed earlier.
Looking at the situation from a different point of view, upper limits
on the wind absorption are above $\taustar \approx 0.5$ for all but
one line complex in the spectrum.  The unmistakable conclusion is that
the \chandra\ spectrum of \zori\ is consistent with a moderate amount
of wind absorption (as well as the expected degree of broadening from
an embedded wind source), and that at least some wind attenuation is
demanded by the data.

The derived \Rmin\ and $q$ values are consistent with the numerical
simulations of the line-force instability wind shocks, inferred from
simulation output shown in various figures in
\citet{Cooper1996,Cohen1996,Feldmeier1997b,or2002}.
These trends are also qualitatively understood from a theoretical
point of view.  The strong, relatively symmetric diffuse (scattered)
radiation field near the photosphere inhibits the line-force
instability and thus the formation of strong shocks near the
photosphere, and the filling factor is not strongly dependent on
radius because although the propensity of shocks to form eventually
falls off with distance from the photosphere, the cooling timescale
for shock-heated plasma increases with distance.

Given the spatial distribution of hot plasma derived from the
line-profile fits, the continuum attenuation by the overlying cool
wind is governed by the mass-loss rate and wind opacity.  In the model
we have employed, the overall wind attenuation is characterized by the
optical depth parameter, 
\[
\taustar \equiv \frac{\kappa\mdot}{4\pi\vinf\Rstar}.
\]
Using mean values from Table
\ref{tab:zori_params} and a wind opacity value of $\kappa \approx 125$
cm$^2$ g$^{-1}$, we expect $\taustar \approx 3$.  The value for the
wind opacity is taken from fig.\ 4 in \citet{Cohen1996}.  Of all the
values that go into this calculation, the mass-loss rate is probably
the most uncertain, followed by the wind opacity and the star's
radius.  The terminal velocity is probably known to within ten or
twenty percent (which is the range of values found in the literature).

Thus, the value of the wind optical depth parametrized by \taustar, as
derived from the observed X-ray line profiles, is about an order of
magnitude lower than the expected value.  This is similar to what is
seen in \zpup\ \citep{kco2003}, where the observed value of \taustar\
is almost an order of magnitude lower than expected (there is also a
fair amount of uncertainty in the relevant properties of \zpup).  The
expected \taustar\ value for \zpup\ is about a factor of two larger
than that for \zori, primarily because of the earlier type star's
larger mass-loss rate.

The fact that the X-ray line profiles of \zpup, and now \zori,
indicate lower than expected wind optical depths is consistent with
recent work that suggests that O star mass-loss rates may have been
overestimated by a factor of three or more, and perhaps up to an order
of magnitude \citep{Bouret2005} due to clumping (which affects density
squared mass-loss diagnostics, such as radio free-free and H${\alpha}$
emission). This result is not inconsistent with the traditional UV
absorption-line based mass-loss rate estimates of hot-star winds,
which have always been subject to uncertainty due to the difficulty of
reliably accounting for ionization distribution effects. In fact,
other recent work, focusing on far-UV absorption line studies as
diagnostics of mass loss and wind ionization in many O and B
supergiants, indicates that mass-loss rates based on UV absorption
line analysis may be overestimated by as much as an order of magnitude
\citep*{fmp2006}. 

Recent detailed multi-wavelength modeling of a large sample of O
giants and supergiants (but not including \zori) indicates mass-loss
rate overestimates of at least a factor of two, assuming that the far
wind, where the radio free-free emission arises, is unclumped, and
more than a factor of two if the far wind is significantly clumped
\citep{Puls2006}. This work also shows that there are star-to-star
variations in clumping factors and a somewhat strong radial dependence
- at least for some stars - to the clumping factor.  We note in this
context that the X-ray profiles will in general be most sensitive to
clumping in the region near and just above \Rmin\ (so, $R \approx
2\Rstar$, which is ``region 2'' in the analysis of \citet{Puls2006}).
These various threads of evidence for lower mass-loss rates are also
consistent with the energy budget analysis of wind-blown bubbles and
superbubbles (see, e.g., \citet{Naze2002,Cooper2004} and references
therein; but see also \citet*{fhy2006} for the role played by the
swept-up wind from earlier evolutionary stages in Wolf-Rayet bubbles).

Even apart from its effect on mass-loss rate estimates, clumping
itself has the potential to reduce the effective opacity of a stellar
wind \citep*{foh2003,ofh2004}.  This effect might more accurately be
termed ``porosity,'' as it presumes the existence of a low-density
interclump channels that can potentially allow photons to escape the
wind more easily \citep*{ogs2004}.
\citet*{ofh2005,ofh2006} have recently computed X-ray line profiles
for specific models incorporating geometrically thin, radially
compressed shells, comparing their results with observed profiles for
several hot stars (including \zori).  Within the assumptions in their
model, these authors show that a very optically thick emission line,
with a significant skewness in a smooth wind, can be made moderately
more symmetric with an interclump spacing of 0.2 \Rstar, and can be
made nearly symmetric with an interclump spacing of 2 \Rstar\ (see
fig.\ 1 in \citet{ofh2005}).
Similarly, using a parametrized model of isotropic clumping, based
generally on the porosity formalism introduced by \citet{ogs2004},
\citet{OC06} find that obtaining symmetric X-ray emission profiles
from an otherwise optically thick wind requires a quite large
``porosity length'' $h \equiv \ell/f$, where $\ell$ represents the
characteristic clump scale, and $f$ if the clump volume filling
factor.  Specifically it requires $h$ of order the local radius $r$.

In this context, it is thus important to stress that while the
mass-loss overestimates due to clumping depend only on the density
contrast between the clumps and the interclump medium (and thus the
volume filling factor), for porosity to affect the line profiles
directly, the density contrast must be accompanied by a large clump
scale, or interclump spacing.  The most sophisticated numerical
treatment of the line-force instability shows structure on small
($\ell \ll\Rstar$) spatial scales, with only moderately compressed
volume filling factors ($f \approx 0.1$ \citep{do2003}).

These results make it difficult to see how the wind inhomogeneities
produced by the instability and which, presumably, are directly
related to the shock-heating responsible for the X-ray emission
itself, could lead to a significant porosity effect on the X-ray line
profiles.  In light of the several independent lines of evidence for
lower O star wind mass-loss rates, we suspect that lower wind column
densities are the cause of the order of magnitude discrepancy between
the \taustar\ values we derive in this paper from fits to the emission
lines in the \chandra\ spectrum of \zori\ and the similar results
derived by \citet{kco2003} for \zpup.  Clumping and the associated
porosity may play some role, but for that role to be significant, the
porosity length in O star winds must be large - of order the local
radius - and the combination of these two effects must reduce the
effective wind optical depth by an order of magnitude.

The results from our X-ray emission line profile analysis should be
consistent with other aspects of the \chandra\ observations.  The
emission measure and temperature information derived from the
observations \citep{cs1983,bsc1996,wc2001} are typical for O
supergiants and do not provide any significant constraints on the
interpretation of the line profiles, aside from simply being broadly
consistent with the expectations of the standard wind-shock scenario.
The most constraining specific X-ray diagnostic in conjunction with
the emission line profiles is the forbidden-to-intercombination
emission line ratio in the helium-like isoelectronic sequence
\citep*{gj1969,bdt1972}.  In the presence of a strong UV field which
can drive photoexcitation of electrons from the upper level of the
forbidden line to the upper level of the intercombination line (2s
$^3S_1$ - 2p $^3P_{1,2}$) and thus reduce the $f/i$ line ratio, it can
be used as a diagnostic of the UV mean intensity and thus of the
distance of the X-ray emitting plasma from the photosphere.

The initial work on the several helium-like $f/i$ ratios seen in the
\chandra\ spectra from \zori\ showed that most of the helium-like ions
were far from the photosphere, consistent with those ions being
embedded in the stellar wind, but that the Si\,{\sc xiii} $f/i$ ratio
implied a location only slightly above the photosphere, which would
generally be considered too close to the star to be consistent with
any wind-shock scenario \citep{wc2001}.  However, a recent reanalysis
of these same data showed that all the helium like ions, including
Si\,{\sc xiii}, are consistent with an onset radius (\Rmin) of about
1.5\Rstar\ \citep{Leutenegger2006}.  This result is, of course,
completely consistent with those we report here for the emission line
profiles of nine other lines in the \chandra\ data.

We can also consider trends in the derived wind profile parameters
within our dataset.  One might expect different lines to have
different morphologies and thus different model parameters either
because different ions form at different temperatures and thus sample
different shocked regions or because lines at different wavelengths
have differing amounts of wind attenuation due to the wavelength
dependence of the opacity of the bulk, cold wind. Regarding the first
possible effect, we note that numerical simulations show a relatively
constant rms velocity dispersion with radius, once shocks begin to
form \citep{ro2002}. Figure 5 in \citet{ro2002} shows, in detail, a
very rapid rise in the velocity dispersion, followed by a very shallow
fall-off with radius.

Regarding the second effect; that of wavelength-dependent attenuation,
photoionization cross sections of cosmically abundant plasma do have a
strong wavelength dependence over a large range of wavelengths.
However, this effect is more complex when the plasma is ionized, as it
is even in the ``cold'' component of a hot-star wind.  Furthermore,
the lines we analyze in this paper span only a factor of two in
wavelength.  Looking at the wind opacity in fig.\ 4 of
\citet{Cohen1996}, we can see that the values of the wind opacity
range only over about a factor of 2 from 600 eV (roughly the photon
energy of the O\,{\sc vii} lines near 22 \AA\ which are the
longest-wavelength lines to which we fit the wind profile model) to
1000 eV (roughly the photon energy of the Ne\,{\sc x} \lya\ line,
which, at $\lambda=12.134$ \AA, is the shortest wavelength line we
discuss here).  The variations in the wind opacity on this relatively
small wavelength range are complex and not monotonic because of the
dominance of photoionization edges of oxygen (O$^{+3}$ through
O$^{+5}$).  The appearance of these edges breaks up the usual $E^{-3}$
fall off in opacity, and over this relatively small wavelength range,
makes the opacity roughly constant.  If anything, the longest
wavelength lines in our data (the O\,{\sc vii} lines near 22 \AA) are
subject to less attenuation than the shorter wavelength lines, by
virtue of their being longward of the oxygen K-shell edges (and, in
fact, this emission feature has the lowest upper limit to the
\taustar\ parameter of any of the lines we fit).  In any case, there
are no significant systematic trends in any of the three wind profile
model parameters.  As we discussed above, a single value of each
parameter is consistent with all the data.  So, although higher
signal-to-noise data in the future may reveal a significant trend,
none is seen in these data.  We should point out, though, that
\citet{ofh2006} noted that radiation transport through a medium with
completely optically thick clumps will not only reduce the effective
wind opacity, but will make the opacity effectively gray. Interpreting
the wavelength dependence of line profile morphologies -- or lack
thereof -- however, requires both a detailed evaluation of the
wavelength-dependent atomic opacity and its uncertainty, and also
statistical fitting of whatever line profile model may be appropriate
along with formal constraints on confidence limits of the parameters
of that model.

Finally, we note that each line or line complex is well fit, in a
statistical sense, by the relatively simple, spherically symmetric
wind-profile model we employ here.  Future higher resolution and/or
higher signal-to-noise spectra could show evidence for signatures of
wind asymmetry or of time variability in the line profiles (perhaps
much like DACs see in UV absorption lines from the winds of hot stars
or like moving emission bumps seen in WR spectra).  There is, however,
no need at this point to invoke either of these effects nor any others
that go beyond the basic model we have used here.

\section{Conclusions}

The fundamental observational conclusions of this work are that the
X-ray emission lines of the late O supergiant \zori\ are broad, blue
shifted, and modestly asymmetric, which is qualitatively consistent
with the general picture of hot, X-ray emitting plasma embedded in an
expanding, spherically symmetric stellar wind.  These results come
both from fitting a physics-based empirical wind-profile model to nine
emission lines in the \chandra\ MEG spectrum, and also from attempts
to fit Gaussian line-profile models and a non-parametric analysis of
the line shapes via the calculation of the first three moments of
seven unblended lines.

There is no need, based on the observed line profiles, to invoke {\it
  ad hoc} coronal emission or other non-standard X-ray production
mechanisms.  However, the amount of attenuation by the bulk, cold
stellar wind is significantly less than would be expected by a simple
application of the assumed mass-loss rate, standard warm plasma
opacities, and the assumption of a spherically symmetric, smooth
stellar wind.  Qualitatively, this result is consistent with the
results of a similar analysis of the \chandra\ spectrum of the early O
star, \zpup\ \citep{kco2003}. And the smaller-than-expected wind
attenuation leaves an observational signature that explains why
previous studies, in which Gaussian profiles were fit and then
analyzed ``by eye,'' did not identify the signature of wind
attenuation.  The emission lines, though significantly blue shifted,
are only modestly asymmetric, and in fact, any individual line can be
at least marginally fit by a blue shifted Gaussian.  For the strongest
lines, however, there is a significant improvement in the fits based
on the wind-profile models as compared to those based on Gaussians.

These results, taken together with the earlier ones on the X-ray line
profiles of \zpup, indicate then that the standard wind-shock scenario
is adequate for explaining the high-resolution X-ray spectra for
normal O supergiants.  Unusual hot stars, such as \tori\ and \tsco, do
not fit into this paradigm, perhaps because of their extreme youth,
but there is no reason, especially now that the helium-like $f/i$ line
ratios have also been reanalyzed \citep{Leutenegger2006}, to suppose
that all hot stars, with the sole exception of \zpup, pose an
insurmountable challenge to the wind-shock model of X-ray production.
That being said, the wind-shock model still has various difficulties
in accounting in detail for the observed trends in X-ray properties
among OB stars, and there are many open questions about the specific
ingredients of a correct wind-shock model.  But the nature of X-ray
emission line profiles in O supergiants, while providing some
interesting constraints and presenting a puzzle about wind optical
depths, does not require us to completely discard the wind-shock
paradigm or lead us to invoke coronal models for explaining hot-star
X-ray emission.  The lower than expected wind optical depths derived
from the X-ray line profiles do, however, add to the debate about O
star mass-loss rates and the role of wind clumping.

\section*{Acknowledgments}

We wish to thank J. Elliot Reed for initial work with the {\it
  Mathematica} modeling code and the moment analysis. DHC acknowledges
NASA contract AR5-6003X to Swarthmore College through the Chandra
X-ray Center. SPO acknowledges NSF grants AST-0097983 and AST-0507581.
RHK acknowledges NASA grant NAG5-9461 to Prism Computational Sciences,
and also the support of the Howard Hughes Medical Institute grant to
Swarthmore College.  DHC and KG thank the National Science Foundation
for its support to the Keck Northeast Astronomy Consortium through
grant AST-0353997.




\begin{table*}
 \centering
\begin{minipage}{160mm}
\caption{Stellar properties of $\zeta$ Orionis from the literature.}
\begin{tabular}{@{}lcccccc@{}}
\hline
Reference & $M$ & $R$ & $M_{\rm v}$ & $B-V$ & $\dot{M}$ & $v_{\infty}$  
\\
 & $(\rm M_{\odot})$ & $(\rm R_{\odot})$ &  &  & $(10^{-6} \rm M_{\odot}\ yr^{-1})$ & $(\rm km\ s^{-1})$ 
\\   
\hline

Lamers \& Leitherer (1993) & 49 & 31 & -7.0 & --- & 2.51 & 2100 \\
Prinja, Barlow, \& Howarth (1990) &--- &--- & --- & --- & --- & 1860 \\
Blomme (1990)& 37 &--- & -6.7 & --- & --- & 2400 \\
Groenewegen et al. (1989)& 41 & 26 & -6.6 & --- & --- & 2100 \\
Voels et al. (1989)&34 &24 & --- & -0.27 & --- & --- \\
Wilson and Dopita (1985)& 25 & 20 & --- & --- & 1.58 & 2190 \\

\hline
\end{tabular}
\label{tab:zori_params}
\end{minipage}
\end{table*}


\begin{table*}
 \centering
 \begin{minipage}{160mm}
\caption{Gaussian line profile fits to the emission lines.}
\begin{tabular}{@{}lccccc@{}}
\hline
Ion & $\lambda_{\rm o}$ & Centroid & HWHM \\
 & (\AA) & (km s$^{-1}$) & (km s$^{-1}$) \\
\hline

N\,{\sc vii} & 24.781 & $-110 \pm 140$ & $1380_{-130}^{+140}$ \\
O\,{\sc viii} & 18.969 & $-150 \pm 30$ & $810 \pm 30$  \\
O\,{\sc vii} & 18.627 & $-380_{-80}^{+90}$ & $500_{-70}^{+120}$  \\
O\,{\sc viii} & 16.006 & $-100_{-80}^{+60}$ & $880 \pm 70$  \\
Fe\,{\sc xvii} & 15.014 & $-180_{-50}^{+40}$ & $830_{-40}^{+50}$ \\
Ne\,{\sc x} & 12.134 & $-150 \pm 50$ & $980 \pm 50$ \\
Ne\,{\sc ix} & 11.544 & $-390 \pm 140$ & $1360_{-150}^{+160}$ \\

\hline
\end{tabular}
\label{tab:gauss_fits}
\end{minipage}
\end{table*}


\begin{table*}
 \centering
 \begin{minipage}{160mm}
\caption{First and third moments of the emission line profiles.}
\begin{tabular}{@{}lccccc@{}}
\hline
Ion & $\lambda_{\rm o}$ (\AA) & M1 & M1/uncert. & M3 & M3/uncert. \\
\hline

O\,{\sc viii} & $18.969$ & $-0.0818 \pm 0.0135$ & $-6.08$ &
 $7.9914 \pm 4.8511$ & $1.65$ \\ 
O\,{\sc vii} & $18.627$ &  $-0.1138 \pm 0.0416$ & $-2.74$ &
 $4.5943 \pm 2.9983$ & $1.53$ \\
Fe\,{\sc xvii} & $16.780$ & $-0.1652 \pm 0.0229$ & $-7.32$ &
 $7.5967 \pm 3.9259$ & $1.94$ \\
O\,{\sc viii} & $16.006$ & $-0.0464 \pm 0.0247$ & $-1.88$ &
 $3.9309 \pm 3.8114$ & $1.03$ \\
Fe\,{\sc xvii} & $15.014$ & $-0.0792 \pm 0.0173$ & $-4.58$ &
 $12.5198 \pm 5.5363$ & $2.26$ \\
Ne\,{\sc x} & $12.134$ & $-0.0801 \pm .0194$ & $-4.13$ & $10.1529 \pm 5.4936$ & $1.85$ \\
Ne\,{\sc ix} & $11.544$ & $-0.1108 \pm 0.0368$ & $-3.01$ &
 $2.8778 \pm 4.0377$ & $0.71$ \\

\hline
\end{tabular}
\\Note: M1 and M3 are the first and third moments of the line
profiles, respectively.  The following columns show the ratio of the
values of these moments, for the indicated unblended lines, to their
formal uncertainties.  We interpret the values in these columns as
significance indicators of the first and third moments' deviation from
zero, as described in Sec.\ 3.
\label{tab:moments}
\end{minipage}
\end{table*}


\begin{table*}
 \begin{minipage}{160mm}
\caption{Wind profile model parameters fit to the data.}
\begin{tabular}{@{}lccccc@{}}
\hline
Ion & $\lambda_{\rm o}$ (\AA) & $q$ & $R_{\rm min}/R_{\rm \ast}$ & $\tau_{\ast}$ & Goodness of fit$^{a}$ \\
\hline

O\,{\sc vii} & 21.804, 21.602 & 
 $-0.30^{+.27}_{-.19}$  &
 $1.66^{+.15}_{-.13}$ & 
 $0.06^{+.14}_{-.06}$ &
  0.33 \\
O\,{\sc viii} & 18.969 &
 $-0.12^{+.29}_{-.22}$ &
 $1.61^{+.14}_{-.12}$ &
 $0.26^{+.20}_{-.13}$ &
  0.67  \\ 
O\,{\sc vii} & 18.627 &
 $0.39^{+1.38}_{-.71}$ &
 $1.29^{+.29}_{-.18}$ &
 $1.34^{+1.86}_{-.74}$ &
  0.49 \\
Fe\,{\sc xvii} & 17.051, 17.096 &
 $-0.41^{+.29}_{-.19}$ &
 $1.28^{+.21}_{-.13}$ &
 $0.76^{+.52}_{-.33}$ &
  0.40  \\
O\,{\sc viii} & 16.006 &
 $-0.41^{+.41}_{-.31}$ &
 $1.51^{+.98}_{-.25}$ &
 $0.27^{+.48}_{-.19}$ &
  0.53 \\
Fe\,{\sc xvii} & 15.014 &
 $-0.47^{+.22}_{-.16}$ & 
 $1.37^{+.15}_{-.14}$ &
 $0.58^{+.41}_{-.25}$ &
  0.29  \\
Ne\,{\sc x} & 12.134 &
 $-0.50^{+.35}_{-.21}$ & 
 $1.55^{+.32}_{-.21}$ &
 $0.45^{+.46}_{-.29}$ &
  0.14 \\
\hline
\end{tabular}
\\$^{a}$Fraction of Monte Carlo simulated datasets that gave
  a C statistic as good or better than that given by the best-fit
  model and the data.  This can be interpreted as a rejection
  probability. Lower values indicate better fits.
\label{tab:fit_params}
\end{minipage}
\end{table*}


\newpage


\newpage


\begin{thebibliography}{99}

\bibitem[\protect\citeauthoryear{Babel \& Montmerle}{1997a}]{bm1997a}
  Babel J., Montmerle T., 1997a, A\&A, 323, 121
   
\bibitem[\protect\citeauthoryear{Babel \& Montmerle}{1997b}]{bm1997b}
  Babel J., Montmerle T., 1997b, ApJ, 485, L29

\bibitem[\protect\citeauthoryear{Berghoefer, Schmitt, \&
    Cassinelli}{Berghoefer et al.}{1996}]{bsc1996} Berghoefer T.W.,
  Schmitt, J.H.M.M., Cassinelli, J.P., 1996, A\&AS, 118, 481

\bibitem[\protect\citeauthoryear{Blomme}{1990}]{Blomme1990} Blomme R.,
  1990, A\&A, 229, 513

\bibitem[\protect\citeauthoryear{Blumenthal, Drake, \&
    Tucker}{Blumenthal et al.}{1972}]{bdt1972} Blumenthal G.R., Drake
  G.W.F., Tucker W.H., 1972, \apj, 172, 205

\bibitem[\protect\citeauthoryear{Bouret, Lanz, \& Hillier}{Bouret et
    al.}{2005}]{Bouret2005} Bouret J.-C., Lanz T., Hillier D.J., 2005,
  \aap, 438, 301

\bibitem[\protect\citeauthoryear{Cash}{1979}]{Cash1979} Cash W., 1979, \apj, 228, 939

\bibitem[\protect\citeauthoryear{Cassinelli \& Olson}{1979}]{co1979}
  Cassinelli J.P., Olson G.L., 1979, \apj, 229, 304

\bibitem[\protect\citeauthoryear{Cassinelli \& Swank}{1983}]{cs1983}
  Cassinelli J.P., Swank J.H., 1983, \apj, 271, 681

\bibitem[\protect\citeauthoryear{Cassinelli et
    al.}{1994}]{Cassinelli1994} Cassinelli J.P., Cohen D.H.,
  MacFarlane J.J., Sanders, W.T., Welsh, B.Y., 1994, \apj, 421, 705

\bibitem[\protect\citeauthoryear{Cassinelli et
    al.}{2001}]{Cassinelli2001} Cassinelli J.P., Miller N.A., Waldron
  W.L., MacFarlane, J.J., Cohen, D.H., 2001, \apjl, 554, L55

\bibitem[\protect\citeauthoryear{Castor, Abbott, \&
    Klein}{1975}]{cak1975} Castor J.I., Abbott D.C., Klein R.I., 1975,
  \apj, 195, 157

\bibitem[\protect\citeauthoryear{Chen \& White}{1991}]{cw1991} Chen
  W., White R.L., 1991, \apj, 366, 512
  
\bibitem[\protect\citeauthoryear{Cohen et al.}{1996}]{Cohen1996} Cohen
  D.H., Cooper R.G., MacFarlane J.J, Owocki S.P., Cassinelli J.P.,
  Wang P., 1996, \apj, 460, 506

\bibitem[\protect\citeauthoryear{Cohen et al.}{2003}]{Cohen2003} Cohen
  D.H., de Messi{\` e}res G.E., MacFarlane J.J., Miller N.A.,
  Cassinelli J.P., Owocki S.P., Liedahl D.A., 2003, \apj, 586, 495
  
\bibitem[\protect\citeauthoryear{Cooper}{1996}]{Cooper1996} Cooper
  R.G., 1996, Ph.D.\ thesis, University of Delaware

\bibitem[\protect\citeauthoryear{Cooper et al.}{2004}]{Cooper2004}
  Cooper R.L, Guerrero M.A., Chu Y.-H., Chen C.-H., Dunne B.C., 2004,
  \apj, 605, 751

\bibitem[\protect\citeauthoryear{Corcoran et al.}{1993}]{Corcoran1993}
  Corcoran M.F., et al., 1993, \apj, 412, 792

\bibitem[\protect\citeauthoryear{Dessart \& Owocki}{2003}]{do2003}
  Dessart L., Owocki S.P., 2003, \aap, 406, L1

\bibitem[\protect\citeauthoryear{Drew, Hoare, \& Denby}{Drew et
    al.}{1994}]{dhd1994} Drew J.E., Hoare M.G., Denby M., 1994,
  \mnras, 266, 917

\bibitem[\protect\citeauthoryear{Feldmeier et
    al.}{1997a}]{Feldmeier1997a} Feldmeier A., Kudritzki R.--P., Palsa
  R., Pauldrach A.W.A., Puls J., 1997a, \aap, 320, 899

\bibitem[\protect\citeauthoryear{Feldmeier, Puls, \&
    Pauldrach}{Feldmeier et al.}{1997b}]{Feldmeier1997b} Feldmeier A.,
  Puls J., Pauldrach A.W.A., 1997b, \aap, 322, 878

\bibitem[\protect\citeauthoryear{Feldmeier, Oskinova, \&
    Hamann}{Feldmeier et al.}{2003}]{foh2003} Feldmeier A., Oskinova
  L., Hamann W.-R., 2003, \aap, 403, 217

\bibitem[\protect\citeauthoryear{Freyer, Hensler, \& Yorke}{Freyer et
    al.}{2006}]{fhy2006} Freyer T., Hensler G., Yorke H.W., 2006,
  \apj, 638, 262

\bibitem[\protect\citeauthoryear{Fullerton, Massa, \&
    Prinja}{Fullerton et al.}{2006}]{fmp2006} Fullerton A.W., Massa
  D.L., Prinja R.K., 2006, \apj, in press (astro-ph/0510252)

\bibitem[\protect\citeauthoryear{Gabriel \& Jordan}{1969}]{gj1969}
  Gabriel A.H., Jordan C., 1969, \mnras, 145, 241

\bibitem[\protect\citeauthoryear{Gagn\'{e} et al.}{1997}]{Gagne1997}
  Gagn\'{e} M., Caillault J.--P., Stauffer J.R., Linsky J.L., 1997,
  \apj, 478, L87

\bibitem[\protect\citeauthoryear{Gagn\'{e} et al.}{2005}]{Gagne2005}
  Gagn\'{e} M., Oksala M., Cohen D.H., Tonnesen S.K., ud-Doula A.,
  Owocki S.P., Townsend R.H.D., MacFarlane J.J., 2005, \apj, 628, 986

\bibitem[\protect\citeauthoryear{Groenewegen, Lamers, \&
    Pauldrach}{Groenewegen et al.}{1989}]{glp1989} Groenewegen M.A.T.,
  Lamers H.J.G.L.M., Pauldrach A.W.A., 1989, \aap, 221, 78
 
\bibitem[\protect\citeauthoryear{Hillier et al.}{1993}]{Hillier1993}
  Hillier D.J., Kudritzki R.--P., Pauldrach A.W.A., Baade D.,
  Cassinelli J.P., Puls J., Schmitt J.H.M.M., 1993, \apj, 276, 117

\bibitem[\protect\citeauthoryear{Ignace}{2001}]{Ignace2001} Ignace R.,
  2001, \apjl, 549, L119

\bibitem[\protect\citeauthoryear{Kahn et al.}{2001}]{Kahn2001} Kahn
  S.M., Leutenegger M.A., Cottam J., Rauw G., Vreux J.--M., den
  Boggende A.J.F., Mewe R., G\"udel M., 2001, \aap, 365, L312

\bibitem[\protect\citeauthoryear{Kramer, Cohen, \& Owocki}{Kramer et
    al.}{2003}]{kco2003} Kramer R.H., Cohen D.H., Owocki S.P., 2003,
  \apj, 592, 532


\bibitem[\protect\citeauthoryear{Kudritzki et
    al.}{1996}]{Kudritzki1996} Kudritzki R.--P., Palsa R., Feldmeier
  A., Puls J., Pauldrach A.W.A., 1996, in Zimmermann H.U., Tr\"{u}mper
  J., Yorke H., eds., R\"{o}ntgenstrahlung from the Universe. MPE,
  Munich, p.\ 9

\bibitem[\protect\citeauthoryear{Lamers \& Cassinelli}{1999}]{lc1999}
  Lamers H.J.G.L.M., Cassinelli, J.P., 1999, Introduction to Stellar
  Winds. Cambridge Univ. Press, Cambridge, U.K.

\bibitem[\protect\citeauthoryear{Lamers \& Leitherer}{1993}]{ll1993}
  Lamers H.J.G.L.M., Leitherer C.M., 1993, \apj, 412, 771

\bibitem[\protect\citeauthoryear{Leutenegger et
    al.}{2006}]{Leutenegger2006} Leutenegger M., Paerels F., Kahn S.,
  Cohen D.H., 2006, \apj, submitted

\bibitem[\protect\citeauthoryear{Lucy}{1982}]{Lucy1982} Lucy L.B.,
  1982, \apj, 255, 286

\bibitem[\protect\citeauthoryear{Lucy \& Solomon}{1970}]{ls1970} Lucy
  L.B., Solomon P.M., 1970, \apj, 159, 879

\bibitem[\protect\citeauthoryear{Lucy \& White}{1980}]{lw1980} Lucy
  L.B., White R.L., 1980, \apj, 241, 300

\bibitem[\protect\citeauthoryear{MacFarlane \&
    Cassinelli}{1989}]{mc1989} MacFarlane J.J., Cassinelli J.P., 1989,
  \apj, 347, 1090
  
\bibitem[\protect\citeauthoryear{MacFarlane et
    al.}{1991}]{MacFarlane1991} MacFarlane J.J., Cassinelli J.P.,
  Welsh B.Y., Vedder P.W., Vallerga J.V., Waldron W.L., 1991, \apj,
  380, 564

\bibitem[\protect\citeauthoryear{Mauche, Liedahl, \& Fournier}{Mauche
    et al.}{2001}]{mlf01} Mauche C.W., Liedahl D.A., Fournier K.B.,
  2001, \apj, 560, 992

\bibitem[\protect\citeauthoryear{Ma{\'{\i}}z-Apell{\' a}niz et
    al.}{2004}]{Maiz-Apellaniz2004} Ma{\'{\i}}z-Apell{\' a}niz J.,
  Walborn N.R., Galu{\' e} H.{\' A}., Wei L.H., 2004, \apjs, 151, 103

\bibitem[\protect\citeauthoryear{Mewe et al.}{2003}]{Mewe2003} Mewe
  R., Raassen A.J.J., Cassinelli J.P., van der Hucht K.A., Miller
  N.A., G\"{u}del M., 2003, \aap, 398, 203

\bibitem[\protect\citeauthoryear{Miller et al.}{2002}]{Miller2002}
  Miller N.A., Cassinelli J.P., Waldron W.L., MacFarlane J.J., Cohen
  D.H., 2002, \apj, 577, 951

\bibitem[\protect\citeauthoryear{Mullan}{1984}]{Mullan1984} Mullan
  D.J., 1984, \apj, 283, 303

\bibitem[\protect\citeauthoryear{Mullan \& MacDonald}{2005}]{mm2005}
  Mullan D.J., MacDonald J., 2005, \mnras, 356, 1139

\bibitem[\protect\citeauthoryear{Mullan \& Waldron}{2006}]{mw2006}
  Mullan D.J., Waldron W.L., 2006, \apj, 637, 506

\bibitem[\protect\citeauthoryear{Naz\'{e} et al.}{2002}]{Naze2002}
  Naz\'{e} Y., Chu Y.-H., Guerrero M.A., Oey M.S., Gruendl R.A.,
  Smith R.C., 2002, \apj, 124, 3325

\bibitem[\protect\citeauthoryear{Oskinova, Feldmeier, \&
    Hamann}{Oskinova et al.}{2004}]{ofh2004} Oskinova L., Feldmeier
  A., Hamann W.--R., 2004, \aap, 422, 675

\bibitem[\protect\citeauthoryear{Oskinova, Feldmeier, \&
    Hamann}{Oskinova et al.}{2005}]{ofh2005} Oskinova L., Feldmeier
  A., Hamann W.--R., 2005, in The X-ray Universe. El Escorial, Madrid,
  Spain (astro-ph/0511019)

\bibitem[\protect\citeauthoryear{Oskinova, Feldmeier, \&
    Hamann}{Oskinova et al.}{2006}]{ofh2006} Oskinova L., Feldmeier
  A., Hamann W.--R., 2006, \mnras, submitted

\bibitem[\protect\citeauthoryear{Owocki \& Cohen}{1999}]{oc1999}
  Owocki S.P., Cohen D.H., 1999, \apj, 520, 833
   
\bibitem[\protect\citeauthoryear{Owocki \& Cohen}{2001}]{oc2001}
  Owocki S.P., Cohen D.H., 2001, \apj, 559, 1108

\bibitem[\protect\citeauthoryear{Owocki \& Cohen}{2006}]{OC06} Owocki
  S.P., Cohen D.H., 2006, \apj, submitted

\bibitem[\protect\citeauthoryear{Owocki \& Runacres}{2002}]{or2002}
  Owocki S.P., Runacres M.C., 2002, \aap, 318, 1015

\bibitem[\protect\citeauthoryear{Owocki \& Rybicki}{1984}]{or1984}
  Owocki S.P., Rybicki G.B., 1984, \apj, 284, 337

\bibitem[\protect\citeauthoryear{Owocki, Castor, \& Rybicki}{Owocki et
    al.}{1988}]{ocr1988} Owocki S.P., Castor J.I., Rybicki G.B., 1988,
  \apj, 335, 914
 
\bibitem[\protect\citeauthoryear{Owocki, Gayley, \& Shaviv}{Owocki et
    al.}{2004}]{ogs2004} Owocki S.P., Gayley K.G., Shaviv N.J., 2004,
  \apj, 616, 525

\bibitem[\protect\citeauthoryear{Pallavicini et
    al.}{1981}]{Pallavicini1981} Pallavicini R., Golub L., Rosner R.,
  Vaiana G.S., Ayres T., Linsky J.L., 1981, \apj, 248, 279

\bibitem[\protect\citeauthoryear{Perryman et al.}{1997}]{Perryman1997}
  Perryman M.A.C., et al., 1997, \aap, 323, L49

\bibitem[\protect\citeauthoryear{Prinja, Barlow, \& Howarth}{Prinja et
    al.}{1990}]{pbh1990} Prinja R.K., Barlow M.J., Howarth I.D., 1990,
  \apj, 361, 607

\bibitem[\protect\citeauthoryear{Puls et al.}{2006}]{Puls2006} Puls
  J., Markova N., Scuderi S., Stanghellini C., Taranova O.G., Burnley
  A.W., Howarth I.D., 2006, \aap, in press

\bibitem[\protect\citeauthoryear{Runacres \& Owocki}{2002}]{ro2002}
  Runacres M.C., Owocki S.P., 2002, \aap, 381, 1015

\bibitem[\protect\citeauthoryear{Schulz et al.}{2003}]{Schulz2003}
  Schulz N.S., Canizares C., Huenemoerder D., Tibbets K., 2003, \apj,
  595, 365
  
\bibitem[\protect\citeauthoryear{Smith et al.}{1993}]{Smith1993} Smith
  M.A., Grady C.A., Peters G.J., Feigelson E.D., 1993, \apjl, 409, L49

\bibitem[\protect\citeauthoryear{Smith et al.}{2004}]{Smith2004} Smith
  M.A., Cohen D.H., Gu M.F., Robinson R.D., Evans N.R., Schran P.G.,
  2004, \apj, 600, 972

\bibitem[\protect\citeauthoryear{ud-Doula \& Owocki}{2002}]{uo2002}
  ud-Doula A., Owocki S.P., 2002, \apj, 576, 413

\bibitem[\protect\citeauthoryear{Voels et al.}{1989}]{Voels1989} Voels
  S.A., Bohannan B., Abbott D.C., Hummer D.G., 1989, \apj, 340, 1073

\bibitem[\protect\citeauthoryear{Waldron}{1984}]{Waldron1984} Waldron
  W.L., 1984, \apj, 282, 256

\bibitem[\protect\citeauthoryear{Waldron \& Cassinelli}{2001}]{wc2001}
  Waldron W.L., Cassinelli J.P., 2001, \apjl, 548, L45

\bibitem[\protect\citeauthoryear{Wilson \& Dopita}{1985}]{wd1985}
  Wilson I.R.G., Dopita M.A., 1985, \aap, 149, 295

\end{thebibliography}
\end{document}